\documentclass[twocol]{ametsocV6.1}

\usepackage[uncertainty-mode=compact]{siunitx}
\usepackage{booktabs}
\usepackage{bm}
\usepackage{amsmath} 
\usepackage[capitalise]{cleveref} 
\usepackage{amssymb} 
\usepackage{tikz}
\usetikzlibrary{shapes,arrows, arrows.meta, positioning, calc}

\title{Physical Scales Matter: The Role of Receptive Fields and Advection in Satellite-Based Thunderstorm Nowcasting with Convolutional Neural Networks}

\authors{Christoph Metzl,\aff{a}\correspondingauthor{Christoph Metzl, christoph.metzl@dlr.de} 
Kianusch Vahid Yousefnia,\aff{a}
Richard Müller, \aff{b}
Virginia Poli, \aff{c} \aff{d}
Miria Celano, \aff{c}
Tobias Bölle,\aff{a} 
}

\affiliation{
\aff{a}{Deutsches Zentrum für Luft- und Raumfahrt, Institut für Physik der Atmosphäre, Oberpfaffenhofen, Germany}
\aff{b}{German Weather Service, 63067 Offenbach, Germany}
\aff{c}{Arpae Emilia-Romagna, Hydro-Meteo-Climate Service (SIMC), Bologna, Italy}
\aff{d}{Agenzia ItaliaMeteo, Bologna, Italy}
}

\abstract{The focus of nowcasting development is transitioning from physically motivated advection methods to purely data-driven Machine Learning (ML) approaches. Nevertheless, recent work indicates that incorporating advection into the ML value chain has improved skill for radar-based precipitation nowcasts. However, the generality of this approach and the underlying causes remain unexplored. 
This study investigates the generality by probing the approach on satellite-based thunderstorm nowcasts for the first time. Resorting to a scale argument, we then put forth an explanation when and why skill improvements can be expected. In essence, advection guarantees that thunderstorm patterns relevant for nowcasting are contained in the receptive field at long forecast times.
To test our hypotheses, we train ResU-Nets solving segmentation tasks with lightning observations as ground truth. The input of the Baseline Neural Network (BNN) are short time series of multispectral satellite imagery and lightning observations, whereas the Advection-Informed Neural Network (AINN) additionally receives the Lagrangian persistence nowcast of all input channels at the desired forecast time. 
Overall, we find only a minor skill improvement of the AINN over the BNN when considering fully averaged scores. However, assessing skill conditioned on forecast time and advection speed, we demonstrate that our scale argument correctly predicts the onset of skill improvement of the AINN over the BNN after \SI{2}{\hour} forecast time. We confirm that, generally, advection becomes gradually more important with longer forecast times and higher advection speeds.
Our work accentuates the importance of considering and incorporating the underlying physical scales when designing ML-based forecasting models.}

\begin{document}

\maketitle

%
%
%
\statement
This research tests the effectiveness of combining physically motivated algorithms with artificial intelligence for short-term thunderstorm forecasts. Our neural network uses geostationary satellite imagery and lightning observations together with their advection forecasts as inputs for predicting the probability of lightning occurrence. Additionally, we propose a simple scale-based explanation for when and why incorporating physically motivated algorithms benefit convolutional neural networks, and we test this hypothesis. We find that our scale argument can roughly predict the onset of usefulness. The application of the scale argument is quite general and can aid in the development of improved forecasting models in multiple domains.
%
%
%

%

\section{Introduction}\label{sec:introduction}
Extreme weather events related to thunderstorm activity such as lightning, flash floods, hail, and strong winds, are responsible for severe economic loss \citep{hoeppe_trends_2016} and can be a threat to human life \citep{terti_situation-based_2017}. Increases in severity and frequency of thunderstorms driven by climate change are expected \citep{Diffenbaugh2013, Raedler2019, raupach_effects_2021}. This necessitates timely and accurate very short-term forecasts also known as nowcasts \citep{Bojinski2023}. Physically motivated nowcasting methods rely on advection and empirical handcrafted rules derived from experience \citep{Wang2017} and are at present dominant in operational use \citep{james_nowcastmix_2018, pulkkinen_pysteps_2019, pulkkinen_nowcasting_2020, muller_novel_2022}.
Advection requires an estimate of the (large-scale) ambient wind and is typically implemented by solving the optical-flow problem (equation~\eqref{eq:genAdvecEqu} for $s_c = 0$) for a sequence of remote-sensing images \citep{Lucas-1981-15102,bowler_steps_2006,santek2019,apke_comparison_2022}.

The data-driven nature of nowcasting implies ML methods to be a promising alternative. Indeed, the main focus in the development of state-of-the-art nowcasting techniques relies on deep learning specifically \citep{ravuri_skilful_2021, zhang_skilful_2023}. However, it is currently unsettled how to make the best use of ML in this context. Hence, prior work mainly focuses on optimizing the ML workflow as seen in the strong heterogeneity across all relevant parts of the value chain: Considered hazards include precipitation \citep{zhao_advancing_2024}, lightning \citep{geng_deep_2021} and hail \citep{leinonen_thunderstorm_2023}. Ground truth candidates are direct \citep{wang_predrnn_2023}, thresholded \citep{ortland_development_2023} observations or human labels \citep{cintineo_deep-learning_2020}. ML tasks scan classification \citep{vahid_yousefnia_machine-learning_2024}, segmentation \citep{lagerquist_using_2021}, supervised \citep{guo_3d-unet-lstm_2023} and generative regression \citep{leinonen_latent_2023}. Data sources encompass various combinations of radar \citep{trebing_smaat-unet_2021}, satellite \citep{brodehl_end--end_2022} and lightning \citep{zhou_deep_2020} observations as well as weather stations \citep{andrychowicz_deep_2023}, numerical weather prediction \citep{yousefnia_inferring_2024} or digital elevation models \citep{leinonen_seamless_2022}. Strategies to address rare events or class imbalances manifest either in different under-sampling strategies in the dataset composition \citep{ayzel_rainnet_2020} ), in class weighting during model training \citep{bi_nowcasting_2023}, or in other loss function adaptations \citep{yang_customized_2023}. Furthermore, model architectures leverage building blocks of feedforward \citep{yousefnia_increasing_2025}, recurrent \citep{shi_convolutional_2015}, convolutional  \citep{han_convective_2022}, graph neural networks \citep{farahmand_spatialtemporal_2023} and Transformers \citep{bai_rainformer_2022}.

Recent works indicate that including physics-informed domain knowledge in the form of advection directly into the ML value chain is a promising complementary route to improve nowcasting \citep{zhang_skilful_2023, ritvanen_advection-free_2023, ha_deep_2023, pavlik_fully_2024}. All prior work in that direction focuses exclusively on radar-based precipitation nowcasting, but differs in the implementation details. \cite{zhang_skilful_2023} design a separate evolution network tasked with learning the advection, whose outputs are then conditioned upon and refined by a generative adversarial network (GAN). In order to obtain a fully end-to-end optimizable model, they do not utilize traditional optical flow algorithms. 

\cite{ha_deep_2023} also build upon the advection but utilize multiple traditional optical flow algorithms by combining them with a regression model, which in turn serves as an additional input for a U-Net. \cite{ritvanen_advection-free_2023} pursue a separate approach of employing a traditional optical flow algorithm to detach the growth and decay of rainfall
from the advection. They achieve this by training a U-Net to learn the approximated time derivative of the precipitation field in the Lagrangian coordinates. \cite{pavlik_fully_2024} extend the approach of \cite{ritvanen_advection-free_2023}. They keep the U-Net for learning the growth and decay but implement the transformation to Lagrangian coordinates in an end-to-end trainable fashion similar to \cite{zhang_skilful_2023}.

The first aim of this work is to test the generality of the concept of combining advection with ML by shifting the hazard type from precipitation to lightning, the ML task from regression to segmentation, and the input data from radar to satellite and lightning data. While nowcasting based on radar data generally results in higher skill \citep{leinonen_nowcasting_2022}, the proposed formulation of the nowcasting problem has some advantages: Geostationary satellite data are globally available and do not suffer from beam blockage in mountainous orography \citep{mcroberts_detecting_2017}. Furthermore, prior work has shown that, in ML models, satellite imagery in combination with lightning data can be a decent substitute for radar \citep{leinonen_nowcasting_2022}. Learning to segment lightning occurrence probabilistically, this approach also circumvents the observed inability of precipitation based regression tasks to nowcast extreme weather due to blurring \citep{ayzel_rainnet_2020} by directly focusing on the lightning activity associated with strong convective events. One disadvantage of this formulation, however, is that the input field (satellite and lightning data) is different from the output field (pixel-wise probabilities). This impedes recursive implementations as in \cite{ha_deep_2023} and a decoupling of advection from the growth and decay as is implemented by \cite{ritvanen_advection-free_2023} and \cite{pavlik_fully_2024}. Therefore, we opt for training a ResU-Net \citep{zhang_road_2018} for each forecast time to solve a semantic segmentation task. Binarized lightning observations serve as ground truth. To showcase the benefit of combining advection with ML, we train two ML models. The input of the Baseline Neural Network (BNN) are the two latest observations of four channels of the Meteosat Second Generation (MSG) satellite \citep{schmetz_introduction_2002} and aggregated lightning observations. The Advection-Informed Neural Network (AINN) additionally receives the Lagrangian persistence nowcast of all input channels at the desired forecast time. We observe significant average skill improvements of the AINN over the BNN on climatologically consistent test datasets only for forecast times greater than \SI{2}{\hour}. 

This forecast time dependence motivates the second aim of this work: to propose and test a possible explanation for why and under what conditions combining advection with CNNs can lead to improvements in spatiotemporal forecasting problems. Prior work only provides heuristic reasoning pointing to the complex multi-scale nature of the problem \citep{zhang_skilful_2023}. We propose a scale argument relating the underlying physical scales to the side length of the receptive field \citep{Araujo2019}, the intrinsic scale of CNNs. Put simply, the receptive field of any given pixel in the output is the field of vision of a neural network in the input. In particular, this is the only part that can influence the output of that specific pixel. The scale argument formalizes the idea that, for sufficiently high advection speed and forecast time, the part of the remote sensing observations relevant to the nowcast lies outside of the field of vision of the CNN if it does not have access to the Lagrangian persistence nowcast. We test this scale argument by analyzing the relative skill increase of the AINN over the BNN dependent on advection speed and forecast time. The observed onset of significant improvement confirms our prediction derived from the basic underlying scales.

Our results underscore the significance of being aware of the underlying physical scales that govern forecasting problems in general and that even in data-driven approaches these scales need to be addressed sufficiently. Furthermore, we offer a way to gauge the adequacy of CNN architectures for forecasting problems where the underlying scales can be estimated.

The structure of this work is as follows: In \cref{sec:data}, a detailed description of the satellite and the lightning observations is given. \Cref{sec:methods} summarizes the advection algorithm, the semantic segmentation task, our neural network model, the dataset composition strategy and the evaluation metrics. \cref{sec:receptField} formally introduces the concept of the receptive field and the scale argument. In \cref{sec:results}, we report on averaged and advection-speed-conditioned skill metrics and present a case study. \cref{sec:discussion} summarizes our work and elaborates on its implications.

\section{Data}\label{sec:data}

\subsection{Area of Interest}\label{sec:aoe}
\begin{figure}[htbp]
\centering
\includegraphics[width=\columnwidth]{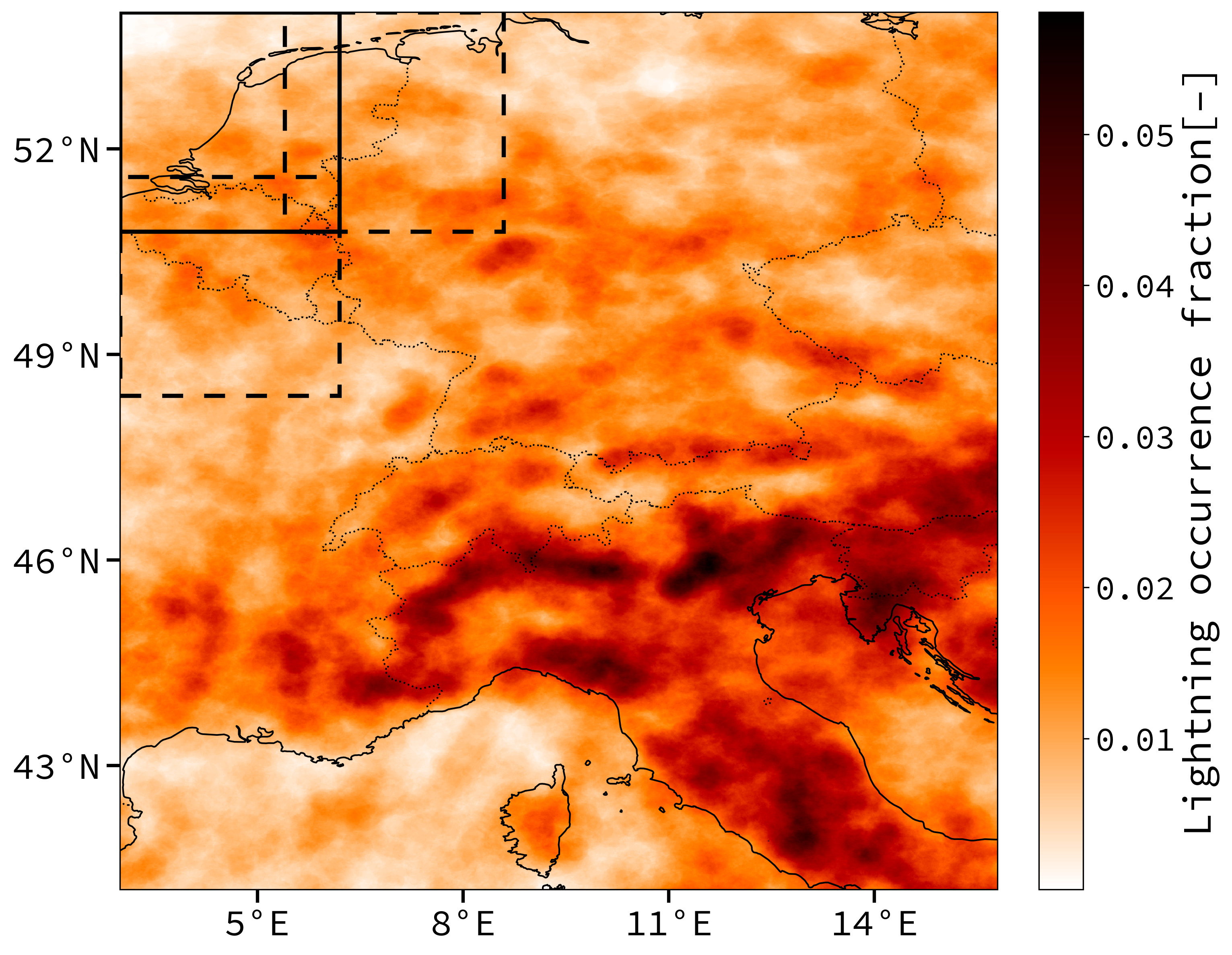}
\caption{Lightning climatology of the area of interest, shown in a Plate Carrée projection. The climatology represents the fraction of time frames in which lightning was detected at each pixel. The lightning observations are from the ground-based lightning detection network LINET (Section \ref{sec:data}\ref{sec:lightObsv}). The spatial distribution is consistent with previous studies \citep{manzato_pan-alpine_2022}. The vertices of the study region are listed counterclockwise from the bottom-left: (\ang{41.2}N, \ang{3}E), (\ang{41.2}N, \ang{15.8}E), (\ang{54}N, \ang{15.8}E), (\ang{54}N, \ang{3}E). Three exemplary crops are overlaid as rectangles with black solid and dashed lines in the upper-left portion of the figure to illustrate the cropping scheme.}
\label{fig:AreaOfInterest}
\end{figure}

The area of interest of this study (\cref{fig:AreaOfInterest}) is roughly centered around the Alps containing among others parts of Germany, France, Italy, Austria and Switzerland. The reason for this choice is the strong thunderstorm activity in this region especially to the South of the Alps \citep{manzato_pan-alpine_2022}. The region, additionally, covers various topographies such as parts of the Mediterranean Sea and the flat Netherlands that complement the mountainous terrain. The geodetic coordinate reference system of choice is WGS84 (EPSG:4326). All data sources are processed on a regular grid with a spacing of \ang{0.0125} ($\approx \SI{1}{\km}$) resulting in $1024 \times 1024$ grid points. The high resolution is chosen to take full advantage of the high resolution visible (HRV) channel introduced in the following (\cref{sec:data}\ref{sec:satObsv}). \cref{fig:AreaOfInterest} displays the area of interest and its lightning climatology calculated from the dataset employed in this study (\cref{sec:data}\ref{sec:lightObsv}) together with an illustration of the cropping scheme. Inspired by \cite{brodehl_end--end_2022}, the area of interest is subdivided into 25 regular crops of size 256 $\times$ 256 that overlap by 64 pixels in both spatial directions to obtain inputs of manageable size for the neural networks.

\subsection{Satellite Observations}\label{sec:satObsv}

Level 1.5 data of the SEVIRI instrument (Spinning Enhanced Visible Infra-Red Imager) onboard the Meteosat Second Generation (MSG) satellite \citep{schmetz_introduction_2002} are used as input for the network. In full disk mode, the SEVIRI instrument scans the Earth every \SI{15}{\minute} with a spatial resolution of \SI{3}{\kilo\meter} at sub-satellite point. This corresponds roughly to a resolution of $\SI{3}{\kilo\meter} \times \SI{5}{\kilo\meter}$ in Central Europe. SEVIRI provides 2 visible (VIS) and 9 infrared (IR) channels for the full disk. The broadband \SIrange{0.4}{1.1}{\micro\meter} high-resolution visible (HRV) channel scans half of the disc with a sub-satellite resolution of \SI{1}{\kilo\meter}. Based on previous studies \citep{mecikalski_forecasting_2006, mecikalski_cloud-top_2010, bedka_objective_2010, leinonen_nowcasting_2022, brodehl_end--end_2022}, only a subset of all channels is used for this work. Besides the HRV channel,  it consists of the infrared window channel with central wavelength \SI{10.80}{\micro\meter} (IR108), the visual window channel with central wavelength \SI{0.81}{\micro\meter} (VIS008) and the water vapor channel  with central wavelength \SI{6.25}{\micro\meter} (WV062). \par
Thunderstorms manifest as rapidly towering, turbulent cumulonimbus clouds and, therefore, appear as cold, optically thick features with considerable local roughness in satellite imagery. This suggests that a combination of cloud temperature and visual roughness information can be used to discriminate thunderstorms from other clouds. In essence, 
the IR108 and WV062 channels provide information about cloud temperatures indicative of rapid vertical growth and overshooting features above the troposphere. Water vapor by itself is further important for thunderstorm dynamics and an important channel for deriving tropospheric motion vectors \citep{schmetz_introduction_2002}. The visual channels indicate optically thick, deep clouds, while regions of strong local gradients (roughness) can be associated with the turbulent nature of thunderstorm updrafts. Using narrow- and broadband visual imagery is expected to combine the discrimination ability with the high local detail. 

Prior work \citep{cintineo_deep-learning_2020} has demonstrated that for applications combining satellite imagery with ground-based observations in ML models, parallax effects exert minimal influence on model performance. Thus, no parallax correction is applied in this study. All channels are resampled onto the grid defined in \cref{sec:data}\ref{sec:aoe} using pytroll \citep{raspaud_pytroll_2018}. All pixel values $\widehat{\phi}_c$ are individually shifted and scaled by the training set channel-wise mean $\mu_c$ and standard deviation $\sigma_c$ before entering the network ($ c \in \{\ \text{HRV}, \text{IR108}, \text{VIS008}, \text{WV062}, \text{LINET}\}$, where LINET abbreviates the lightning detection network discussed in the next section):
\begin{equation}
\phi_c = \frac{\widehat{\phi}_c-\mu_c}{\sigma_c} .
\label{eq:SatScaling}
\end{equation}

\subsection{Lightning Observations}\label{sec:lightObsv}


Lightning observations from the ground-based lightning detection network (LINET) \citep{betz_linetinternational_2009} are used twice in this work. Namely, they constitute a further data source for the input of the network and they also serve as the ground truth of the thunderstorm segmentation task. This is a common choice \citep{Ukkonen2019, geng_deep_2021, leinonen_seamless_2022, vahid_yousefnia_machine-learning_2024} justified by a high and uniform detection efficiency ($\geq\SI{95}{\percent}$ for LINET) and spatial accuracy (\SI{150}{\meter} for LINET). The tabular data is translated onto the grid by the following procedure: For each lightning stroke at position $\bm{x}_{\text{l}}$ and time $t_{\text{l}}$ we increment the number of strokes $n_{\text{LINET}}(\bm{x},t)$ at grid position $\bm{x}$ at time $t$ by 1 if both conditions

\begin{equation}
\left\| \bm{x} - \bm{x}_\text{l} \right\| < \Delta r \quad \text{and}\quad 0 < t-t_\text{l} < \Delta t
\label{eq:thresholds}
\end{equation}
 are satisfied, where $\left\| \cdot \right\|$ denotes the geodesic distance between $\bm{x}$ and $\bm{x}_{\text{l}}$ on the WGS84 ellipsoid. The spatial and temporal thresholds used in this study read $\Delta r = \SI{15}{\kilo\meter}$ and $\Delta t = \SI{15}{\minute}$. The choice of $\Delta t$ reflects the time interval between two satellite images, whereas the value $\Delta r$ has been chosen after consulting previous studies in the literature with similar thresholds \citep{cintineo_probsevere_2022, leinonen_seamless_2022, vahid_yousefnia_machine-learning_2024}. The binary value for the ground truth of the segmentation task is then determined by the condition $n_{\text{LINET}}(\bm{x},t) > 0$. This constitutes the final version for the ground truth of the output of the networks. For the input, further processing is required: Since the number of strokes vary over multiple orders of magnitude and almost \SI{99}{\percent} of all grid points are assigned the value $n_{\text{LINET}}=0$, the distribution of $n_{\text{LINET}}$ cannot be naively shifted and scaled for the lightning input channel using \cref{eq:SatScaling}. Therefore, we first scale $n_{\text{LINET}}$ logarithmically
\begin{equation}
\widehat{\phi}_{\text{LINET}} = \ln{(n_{\text{LINET}}+1)}
\label{eq:logScalingFlashes}
\end{equation}
and then shift and scale $\widehat{\phi}_{\text{LINET}}$ according to \cref{eq:SatScaling}, which is the final form of the lightning input channel of the networks. The additional linear scaling step ensures that the input values are numerically well-conditioned for training, as values closer to zero mean and unit variance are known to improve optimization stability and convergence in deep learning models \citep{Goodfellow-et-al-2016}.

\section{Methods}\label{sec:methods}

In this section, we introduce the three methods to nowcast thunderstorms (as defined by lightning) used in this study. We refer to the hybrid advection-ML model as \emph{Advection-Informed Neural Network} (AINN). For comparison, we use the same ML model without advection, called \emph{Baseline Neural Network} (BNN) and a physically motivated nowcasting, implemented as a \emph{Lagrangian Persistence Nowcast of the Lightning channel} (LPNL). Both ML models yield probabilistic forecasts while LPNL makes categorical predictions.

\subsection{Advection}\label{sec:advection}

The problem of nowcasting is commonly partitioned into an advection and a life-cycle component \citep{bowler_steps_2006, pierce_nowcasting_2012, prudden_review_2020}. Note that existing operational models typically focus on short-term advection, as reliable prediction of full life cycles remains an open challenge. For each channel of the remote sensing observations $\phi_c(\bm{x},t)$ (c.f. \cref{sec:data}) and under the assumption of divergence free advection fields $\bm{v}_c$, this notion is described by the advection equation:

\begin{equation}
\frac{\partial{\phi_c}}{\partial t} + \sum_{j=1}^{2} v_{c,j} \frac{\partial{\phi_c}}{\partial x_j} = s_c ,
\label{eq:genAdvecEqu}
\end{equation}
where $s_c$ is a source-sink term referred to as the life cycle. For short forecast times the left hand side of the equation (the advection of thunderstorm cells along with the large-scale atmospheric motion) is assumed to dominate the error of the nowcast, while the right hand side (internal dynamics) is negligible, implying $s_c = 0$. This reduces the advection equation to a continuity equation. A nowcast performed on this basis is referred to as a Lagrangian persistence nowcast. 

We implement all Lagrangian persistence nowcasts with the pySTEPS library \citep{pulkkinen_pysteps_2019}, which has been shown to work well with satellite data before \citep{burton_satellite-based_2022, smith_evaluating_2024}. The advection fields are estimated based on the default configuration of the dense Lucas–Kanade algorithm \citep{Lucas-1981-15102, bouguet2001pyramidal} using the last two observations. The advection is performed based on the backward interpolate-once semi-Lagrangian extrapolation scheme \citep{germann_scale-dependence_2002,pulkkinen_pysteps_2019}. 
The AINN detailed in \cref{sec:methods}\ref{sec:NN} utilizes the Lagrangian persistence nowcast of all input channels. The advection field is calculated for each channel individually except for the lightning channel. It is advected based on the advection field of the WV062 channel. The reason for this is that optical flow algorithms - including feature-based methods like Lucas–Kanade - tend to struggle when fields become extremely sparse in the sense that they contain only a few isolated detections \citep{muller_novel_2022, leinonen_seamless_2022}.
Lagrangian Persistence Nowcast of the Lightning channel (LPNL) is one of the models evaluated in the results section \cref{sec:results}\ref{sec:skillMetrics}. It represents the class of physically motivated algorithms. 

\begin{figure}[htbp]
\centering
\includegraphics[width=\columnwidth]{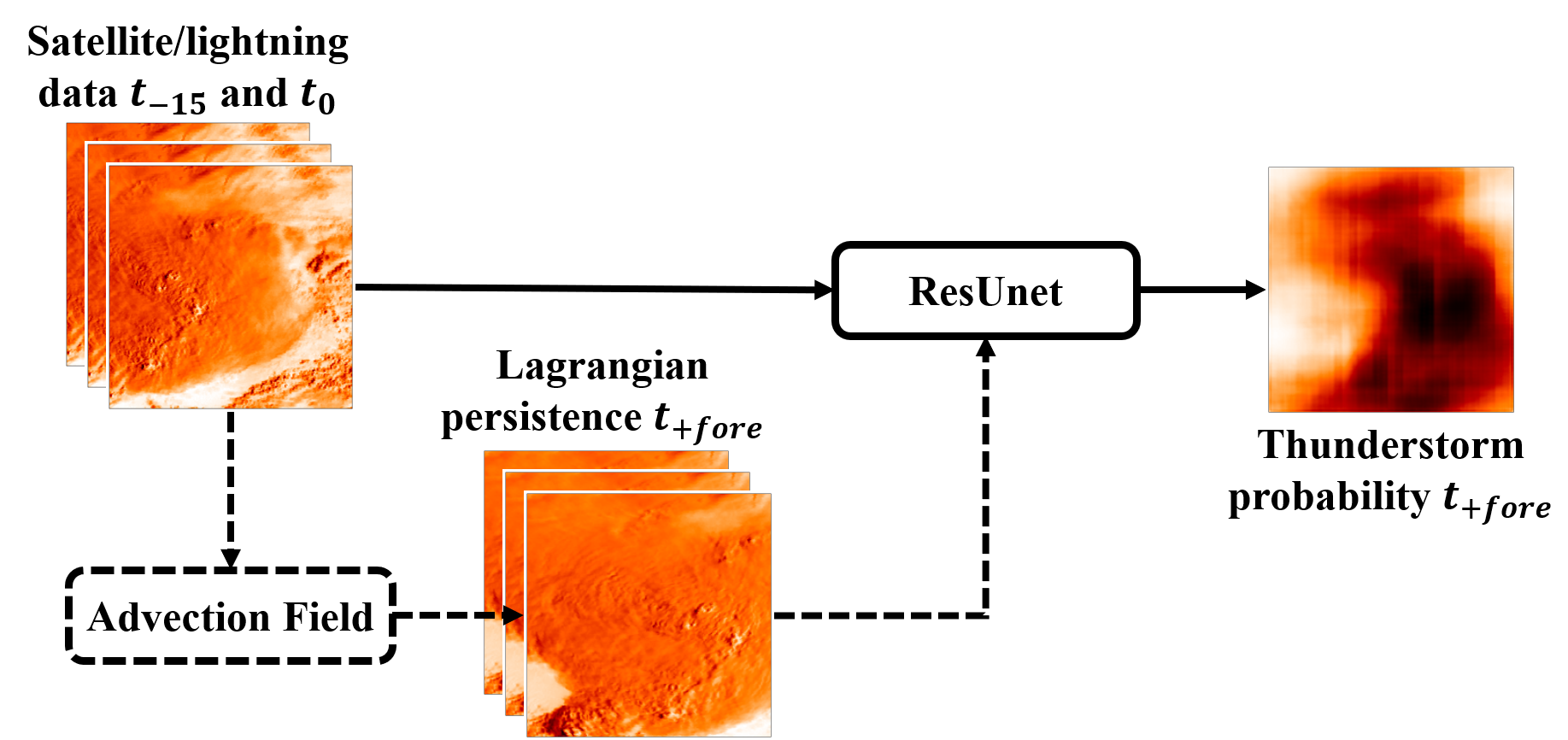}
\caption{Flowchart of the value chain for both the BNN (only solid lines) and the AINN (all lines). The baseline ResU-Net model takes as input the last two observations of four satellite channels and the gridded lightning data to predict a pixel-wise thunderstorm probability for the desired forecast time. The AINN additionally receives the Lagrangian persistence nowcast of each input channel calculated with the advection field derived from the input.}
\label{fig:Flowchart}
\end{figure}

\subsection{Neural Network}\label{sec:NN}

Let $t \in \mathbb{R}_{>0}$ denote the forecast time and $m \in \{\text{BNN, AINN}\}$ the model class. In this study, we train separate models for each forecast time $t \in \{\SI{30}{\minute}, \SI{60}{\minute}, \SI{90}{\minute}, \SI{120}{\minute}, \SI{150}{\minute}, \SI{180}{\minute}\}$. The aim of our neural network models $\bm{f}_{t, m}$ is to forecast the pixel-wise probability of thunderstorm occurrence at $t$ given an array $\bm{X} \in \mathbb{R}^{C \times T_m \times h \times w}$ of satellite and lightning observations. Here $C=5$ is the number of input channels (four satellite and one lightning channel). $T_m$ denotes the number of time steps, which depends on the model class. The Baseline Neural Network (BNN) receives the last two observations, thus $T_{\text{BNN}}=2$. In contrast, the Advection-Informed Neural Network (AINN) additionally receives the Lagrangian persistence nowcast of each channel, thus $T_{\text{AINN}}=3$.  The height and width of the input patch are represented by $h=256$ and $w=256$. Formally, this constitutes a binary segmentation task:

\begin{align}
\bm{f}_{t, m}: \; & \mathbb{R}^{C \times T_m \times h \times w} \rightarrow (0,1)^{h \times w} \\
& \bm{X} \mapsto \bm{f}_{t, m}(\bm{X}; \bm{\theta}_{t, m})
\end{align}
where $\bm{\theta}_{t, m}$ denote the learnable parameters of the models. The number of learnable parameters is the same for all models and equal to 1,633,769. For each forecast time $t$ and model class $m$, a neural network is trained with a data set $D_{t,m}(\text{training}) = \{(\bm{X}_i, \bm{Y}_i)\}_{i=1}^n$ consisting of $n$ tuples. $\bm{Y}_i \in \{0,1\}^{h \times w}$ is the label indicating pixel-wise thunderstorm occurrence (where 1 represents thunderstorm occurrence and 0 represents no thunderstorm occurrence). All datasets are discussed in detail in \cref{sec:methods}\ref{sec:datasetComp}. A flowchart of the value chain for both model classes is illustrated in \cref{fig:Flowchart}.

\begin{figure*}
\centering
\includegraphics[width=\textwidth]{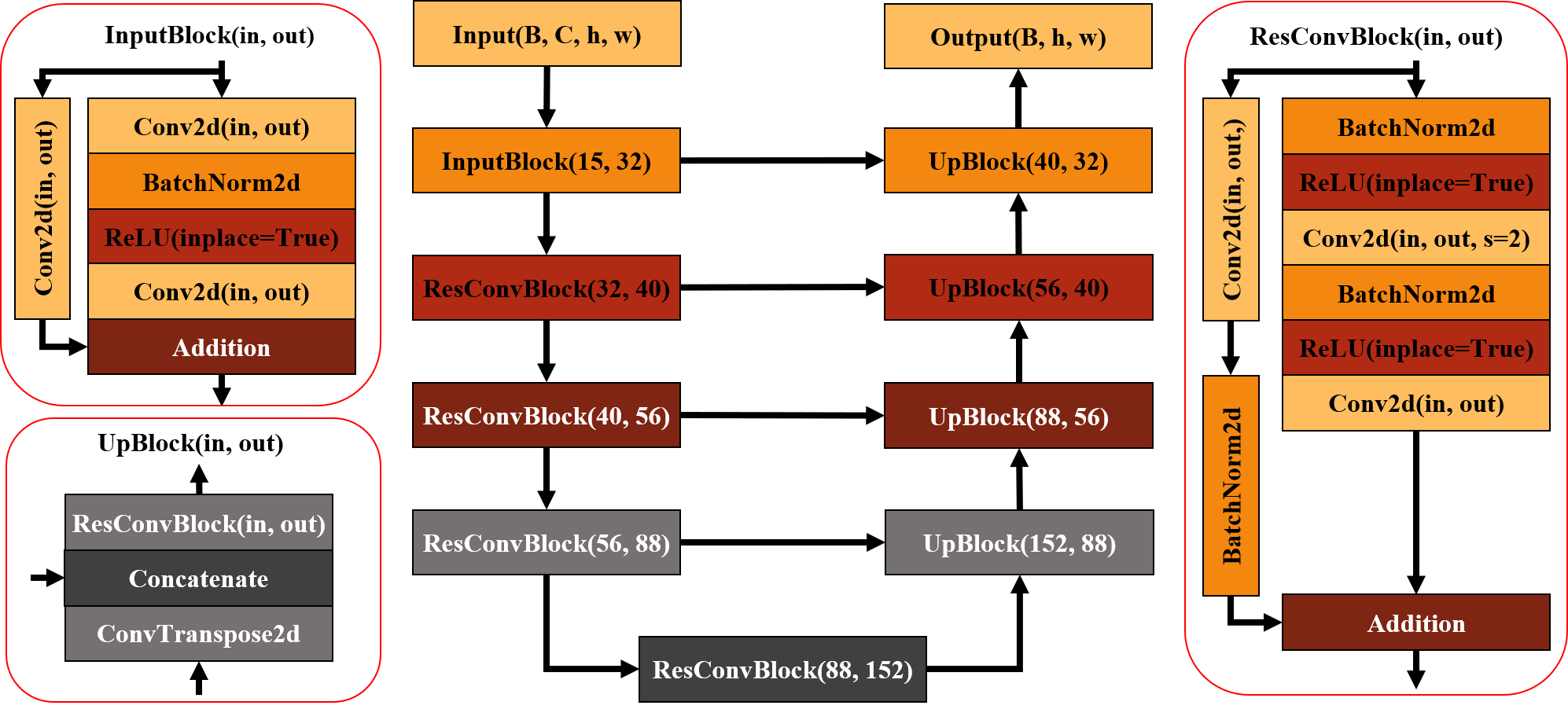}
\caption{Neural network architecture of the ResU-Net of this study. This neural network is used for both the BNN with $(B=128, C=10,h=256,w=256)$ and for the AINN with $(B=128, C=15,h=256,w=256)$. The downsampling is performed with convolutional layers instead of pooling layers to make both down- and upsampling learnable operations.}
\label{fig:NnArchitecture}
\end{figure*}

The U-Net \citep{ronneberger_u-net_2015} and its many variants are still ipso facto the standard model of most image segmentation problems \citep{azad_medical_2024}. It is also heavily featured in ML-based nowcasting models \citep{ayzel_rainnet_2020, cintineo_probsevere_2022, brodehl_end--end_2022, ortland_development_2023} and the architecture chosen for most of the previous work on combining advection with ML \citep{ritvanen_advection-free_2023, ha_deep_2023, pavlik_fully_2024}. Therefore, we also settle for a variant of the U-Net, specifically the ResU-Net \citep{zhang_road_2018}. The specific implementation and architecture hyperparameters are detailed in \cref{fig:NnArchitecture}. We tested multiple versions of the vanilla U-Net and the ResU-Net including 3-dimensional convolutions and enhancements to the bottleneck with multi-head attention and ConvLSTM blocks for better utilization of the information along the time dimension. However, we found that all these modifications underperformed the simpler implementation, where only 2-dimensional convolutions are used and the time dimension is simply absorbed into the channel dimension: $(C=5, T_m, h=256, w=256) \rightarrow (C=5\, T_m, h=256, w=256)$.

The networks are implemented and trained using pytorch \citep{paszke_pytorch_2019}. We optimize the binary cross-entropy with the adam optimizer \citep{Kingma2014} enforcing an L2 regularization and employing the ReduceLROnPlateau learning rate scheduler. All training-specific hyperparameters and their tested variations are documented in \cref{tab:training_hyperparameter}. The validation loss is monitored during training to ensure no overfitting. For each forecast time and model class, the epoch with the smallest validation loss is chosen. The training takes on average \SI{12}{\hour} for the BNN and \SI{14}{\hour} for the AINN on a single NVIDIA HGX A100 80GB 500W GPU.

\subsection{Dataset Composition}\label{sec:datasetComp}

This study utilizes data collected during the summer months from May to October in the years 2018 and 2019. The data is split into three disjoint sets $s \in \{\text{training}, \text{validation}, \text{test}\}$. This is performed similarly to \cite{vahid_yousefnia_machine-learning_2024}: From the available daily samples, we randomly assign 256 days to the training set, while allocating 55 days each to validation and test sets. To minimize information leakage between datasets we define the start of a day to be at 0800 UTC and discard all data where the time of the nowcast (time of the ground truth lightning observations) is between 0800 UTC and 0900 UTC of each day, which we observe to be the hour of minimal lightning activity for our data set. As outlined in \cref{sec:data}\ref{sec:aoe}, the area of interest is subdivided into 25 regular crops of size 256 $\times$ 256 that overlap by 64 pixels in both spatial directions.
This yields three collections of ground truth crops $g(s)=\{\bm{Y}_i\}_{i=1}^{n(s)}$.

For $s \in \{\text{validation}, \text{test}\}$, the final datasets $D_{t,m}(s) = \{(\bm{X}_i, \bm{Y}_i)\}_{i=1}^{n(t,s)}$ are obtained by removing all samples that would require missing or corrupted satellite data as input. This step introduces a dependency on the forecast time and impedes choosing the same forecast-time-independent label set for all models. However, they differ by at most \SI{0.5}{\percent}. The resulting datasets feature a climatologically consistent lightning frequency.

In the case of $s=\text{training}$, all corrupted samples are removed. But, there is a strong class imbalance present in the dataset evidenced by the fact that only \SI{1.15}{\percent} of all pixels are lightning pixels. This is addressed by under-sampling the majority class akin to \cite{cintineo_probsevere_2022}: We calculate the fraction of lightning pixels of each label crop in $g(s)$ and then randomly disregard a portion of the crops with no lightning activity such that the fraction of lightning pixels is equal to $0.05$ for the training set. As a consequence, only \SI{9}{\percent} of training crops contain no lightning pixels at all.

Exemplary for $t=$\SI{180}{\minute}, this approach results in 137,449 crops in the training set,  125,550 in the validation set and 126,125 in the test set.

In particular, the final datasets $D_{t,m}(s)$ are constructed with no special treatment for time of day aside from the buffer hour. Therefore, daytime and nighttime crops are included at a climatologically consistent rate. Visible channels carry no information at night, but the models retain predictive skill. Nonetheless, a reduced skill during nighttime is to be expected \citep{brodehl_end--end_2022}.

\subsection{Evaluation}\label{sec:eval}

The AINN and BNN models output probabilistic forecasts, while the LPNL yields categorical predictions. This fundamental difference necessitates separate evaluation techniques.

Forecasts like the LPNL that output binary predictions are evaluated with skill scores calculated from the 2 $\times$ 2 confusion matrix. It captures the counts of true positives (TP), false positives (FP), false negatives (FN), and true negatives (TN) \citep{Wilks2019-su}. The information dimensionality of the $2 \times 2$ confusion matrix is three. Therefore, it is possible to---in principle---fully characterize the performance of a model with a well chosen triplet of scores \citep{stephenson_use_2000}. One such example is the HBF triplet: the hit rate ($H$), the false alarm rate ($F$), and the bias ratio (B). These metrics are defined as:

\begin{equation}
\begin{aligned}
H = \frac{\text{TP}}{\text{TP} + \text{FN}}, \quad 
F = \frac{\text{FP}}{\text{FP} + \text{TN}}, \quad 
B = \frac{\text{TP} + \text{FP}}{\text{TP} + \text{FN}}
\end{aligned}
\end{equation}
However, for rare events like thunderstorms it is convention to focus on skill scores specifically designed for this situation, such as the Critical Success Index (CSI):

\begin{equation} \text{CSI} = \frac{\text{TP}}{\text{TP} + \text{FP} + \text{FN}} \end{equation}
As the CSI lacks equitability, which means that it does not rate random forecasts and all constant forecasts equally, we choose the Peirce Skill Score (PSS) as the primary categorical skill score for this study. It is both equitable and specifically suited for rare event verification \citep{Wilks2019-su}:

\begin{equation} 
\text{PSS} = H - F
\end{equation}
To compare the LPNL to the neural network models, we threshold the probabilistic outputs such that at each forecast time the CSI is maximized. The thresholds for all models and forecast times are stated in \cref{tab:thresholds_CSI}. The PSS is not suitable for thresholding as the contribution made to it by a true negative and a true positive forecast increases as the event is more or less likely, respectively \cite{Wilks2019-su}.

When comparing the BNN to the AINN we employ evaluation techniques that heed their probabilistic nature. Reliability diagrams \citep{Broecker2007, Wilks2019-su} compare predicted probabilities with observed event frequencies. To that end, we partition the probability range $(0,1)$ into $N_\text{b}$ bins and assign each forecast pixel to its corresponding bin based on its predicted probability. For each bin $i$, we compute the observed event frequency $\overline{o}_i$ and the bin-averaged forecast probability $p_i$. A well-calibrated model exhibits a calibration curve close to the diagonal $\overline{o}_i = p_i$, indicating that predicted probabilities accurately reflect observed frequencies. In addition to the calibration curve, we also examine the distribution of forecast probabilities called the refinement distribution, which provides insight into the model’s ability to distinguish between the positive and negative class. A skillful model not only produces well-calibrated probabilities but also assigns higher probabilities to events more frequently than climatology, demonstrating good resolution. 

While reliability diagrams are conditioned on the forecast probabilities, Precision-Recall (PR) curves are conditioned on the observations. Precision and recall are defined as:

\begin{equation}
\text{Precision} = \frac{\text{TP}}{\text{TP} + \text{FP}}, \quad \text{Recall} = H
\end{equation}

The PR curve is obtained by plotting precision against recall as the threshold of the probabilistic output is systematically varied. A high area under the PR curve (AUC) score indicates a model that maintains high precision and recall across multiple thresholds. These curves are particularly appropriate in the case of strong class imbalance \citep{branco_survey_2016}.

The main skill score of choice for probabilistic models is the Brier Skill Score (BSS). It measures the improvement of the Brier Score (BS) relative to a reference forecast, in our case the climatology $c_t(s)$. The BS is a strictly proper scoring rule \citep{winkler_good_1968}, meaning that it encourages honest probabilistic predictions rather than overconfident or underconfident forecasts and is  essentially the mean squared error of the probability forecasts. With $\|\cdot\|_F$ indicating the Frobenius norm, the two scores are defined as:

\begin{equation}
\text{BS}_{t,m}(s) = \frac{1}{h \, w \, n(t,s)} \sum_{i=1}^{n(t,s)} \|\bm{f}_{t,m}(\bm{X}_i, \bm{\theta}_{t,m}) - \bm{Y}_i\|_F^2
\end{equation}

\begin{equation} \text{BSS}_{t,m}(s) = 1 - \frac{\text{BS}_{t,m}(s)}{\text{BS}_\text{t,m,ref}(s, c_t(s))} \end{equation}

\section{Receptive Field and Scale Argument}\label{sec:receptField}

To our knowledge, there is no detailed explanation for the fact that combining advection algorithms with CNNs can be beneficial. State-of-the-art classical nowcasting methods can determine the advection of cells adequately. They are mostly limited by their inability to model the life cycle \citep{germann_scale-dependence_2002, pierce_nowcasting_2012}. Deep learning models, however, capture some aspects of the life cycle \citep{ayzel_rainnet_2020}. Therefore, if ML models have sufficient capacity, there is no apparent reason why they should lack the ability to learn advective transport, which classical methods capture suitably.

\begin{figure*}
\centering
\includegraphics[width=\textwidth]{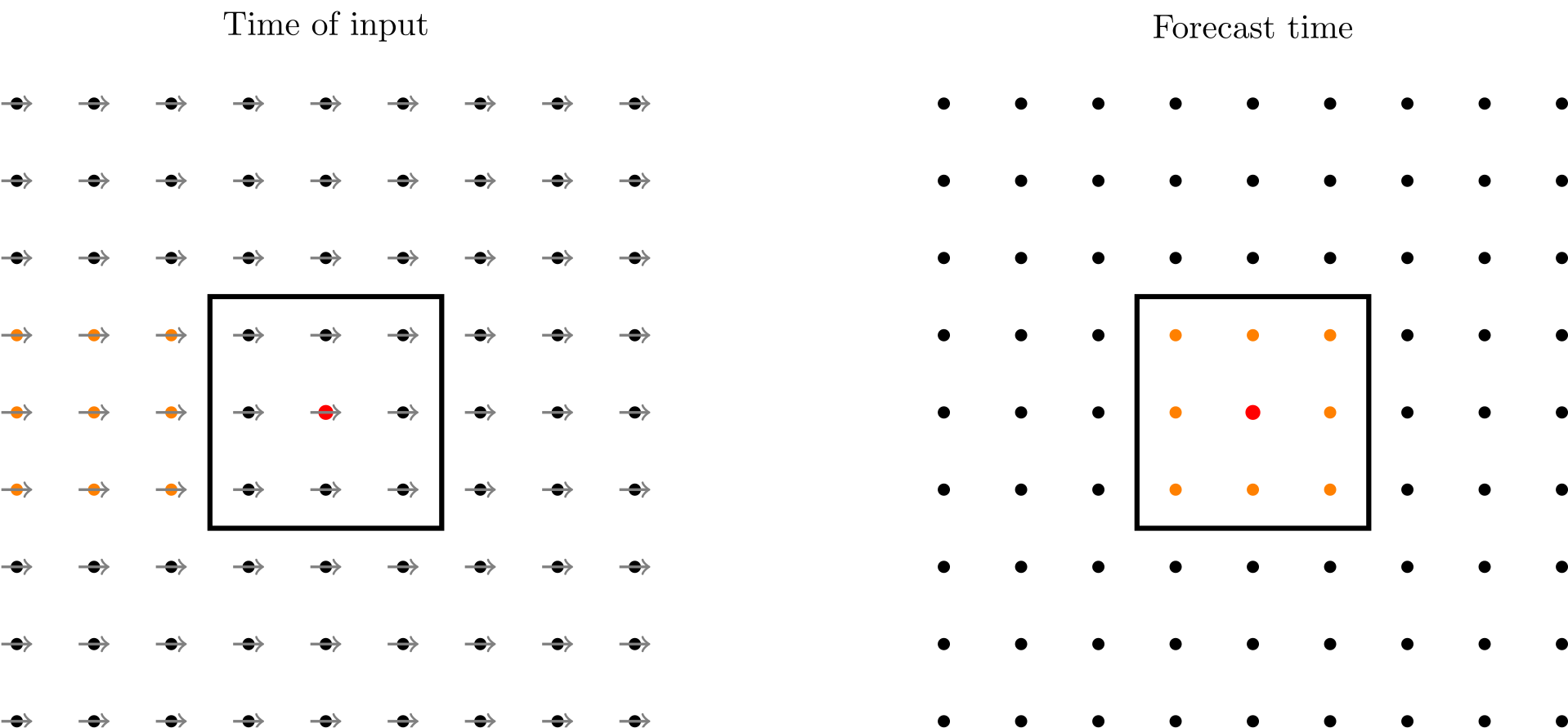}
\caption{Conceptual idea behind the scale argument. For the classification of the central red point the black box represents its receptive field. The orange points indicate the points that actually matter for the classification. At the time of the input (left panel) these points lie outside the receptive field and are advected by the advection field indicated in gray. At forecast time (right panel) the orange points will surround the point of interest. If a neural network does not have access to the advection nowcast it should be unable to make an informed classification attempt.}
\label{fig:receptiveField}
\end{figure*}

We argue that the benefit is connected to the concept of the receptive field of CNNs \citep{Araujo2019}. For each pixel in the output map, the receptive field is the collection of points in the input that can influence that pixel. Therefore, the receptive field can be conceived as the field of vision of a CNN. Put differently, whatever is not in the box at the initial time cannot be utilized by the neural network to infer thunderstorm occurrence at the desired forecast time. For high advection speeds and long forecast times, it is possible that the part of the remote sensing observations that is most relevant to the nowcast of a specific pixel is not within its receptive field as it will be transported there through the advection only at a later time. This idea is illustrated in \cref{fig:receptiveField} and formalized through a scale argument:

For standard CNNs the receptive field is a quadratic box with some side length $r_\text{f}$. Including the Lagrangian persistence nowcast in the input as is the case for the AINN should only lead to significant improvements over the BNN when the spatial scale derived from the large-scale advection speed $v = |\bm{v}_{\text{WV062}}|$ and forecast time $t$ is approximately equal to or greater than $r_\text{f}$:

\begin{equation}
r_\text{f} \gtrsim v \, t .
\label{eq:scaleArgument}
\end{equation}
This implies that the part of the atmosphere that is relevant to forecasting a specific pixel (as this is the part of the atmosphere that will surround the location of the pixel at the desired forecast time) cannot influence the networks decision anymore if it does not have access to the Lagrangian persistence nowcasts obtained from the advection. It is important to stress that this scale argument applies directly only to implementations in which the ML models receive the advection as an additional input as in \cite{zhang_skilful_2023, ha_deep_2023}. It does not, however, apply to implementations that disjoin the advection from the life cycle \citep{ritvanen_advection-free_2023, pavlik_fully_2024}, as they additionally improve the model by forcing it to learn the growth and decay part so there is an additional mechanism at work.

The theoretical maximum side-length of the receptive field $r_\text{f}$ can be computed for CNNs from the following formula:

\begin{equation}
r_\text{f} = \sum_{\ell=1}^{L} \left[ (k_{\ell} - 1) \prod_{j=1}^{\ell-1} s_j \right] + 1
\label{eq:recepField}
\end{equation}
where $L$ is the number of convolutional layers, $k_\ell$ is the filter size of the $\ell$th layer and $s_j$ is the stride of the $j$th layer \citep{Araujo2019}.

For our ResU-Net, $L = 14$, with all convolutional layers using a kernel size $k_\ell=3$. The stride is $s_j=1$ for all layers, except for the four downsampling layers where $s_j=2$. Substituting these values into \cref{eq:recepField} yields a theoretical receptive field of $r_\text{f} = 221$. From this analysis, we conclude that for a forecast time of \SI{2}{\hour} it would require a large-scale advection speed of $v \gtrsim \SI{30}{\meter\per\second}$ for the advection nowcast to benefit the ML model. These are advection speeds that are at the very upper end compared to radiosonde measurements of wind speeds in the troposphere \citep{kruger_influence_2024} (we want to caution here that the scale argument presupposes $v$ to be the large-scale advection speed, which does not necessarily equate to wind speeds directly). Therefore, this implies that the AINN should start to noticeably outperform the BNN on a representative sample only for forecast times greater than \SI{2}{\hour}. We test this in \cref{sec:results}\ref{sec:skillMetrics}. A further implication of this scale argument is that the relative skill increase of the AINN over the BNN should increase with forecast time and advection speed. This is put to the test in \cref{sec:results}\ref{sec:skillAdvectionSpeed}.

\section{Results}\label{sec:results}

First we compare the AINN to the LPNL with categorical skill scores on climatologically consistent test data sets. Then we move on to evaluations of the AINN compared to the BNN that heed their probabilistic characteristics. In the course of this, we perform three types of analysis, which are presented in the following order: Fully averaged skill metrics, conditioned skill metrics, and a case study.

\subsection{Categorical Skill Metrics}\label{sec:categoricalSkillMetrics}
We start by comparing the skill of the AINN to that of the LPNL. The values of the categorical skill scores, which are introduced in  \cref{sec:methods}\ref{sec:eval}, are summarized in \cref{tab:skill_scores}. All values are calculated on the test sets for all forecast times, which are then averaged to have a single score per model. 

\begin{table}[htbp]
\caption{Summary of categorical skill score values for the AINN and the LPNL. The probabilistic predictions of the AINN are converted to binary forecasts. This is done by introducing decision thresholds that maximize the CSI for each forecast time. The thresholds are listed in \cref{tab:thresholds_CSI}. The values are the average over the forecast time of the scores evaluated on the test sets.}
\label{tab:skill_scores}
\centering
\begin{tabular}{ccc}
\toprule
Skill Score &  AINN & LPNL \\
\midrule
PSS & $0.520$ & $ 0.326$ \\
H & $0.529$  & $0.340$  \\
F & $0.010$ & $0.014$ \\
B & $1.147$ & $1.248$ \\
CSI & $0.343$ & $0.191$ \\
\bottomrule
\end{tabular}
\end{table}

\begin{figure}[htbp]
\centering
\includegraphics[width=\columnwidth]{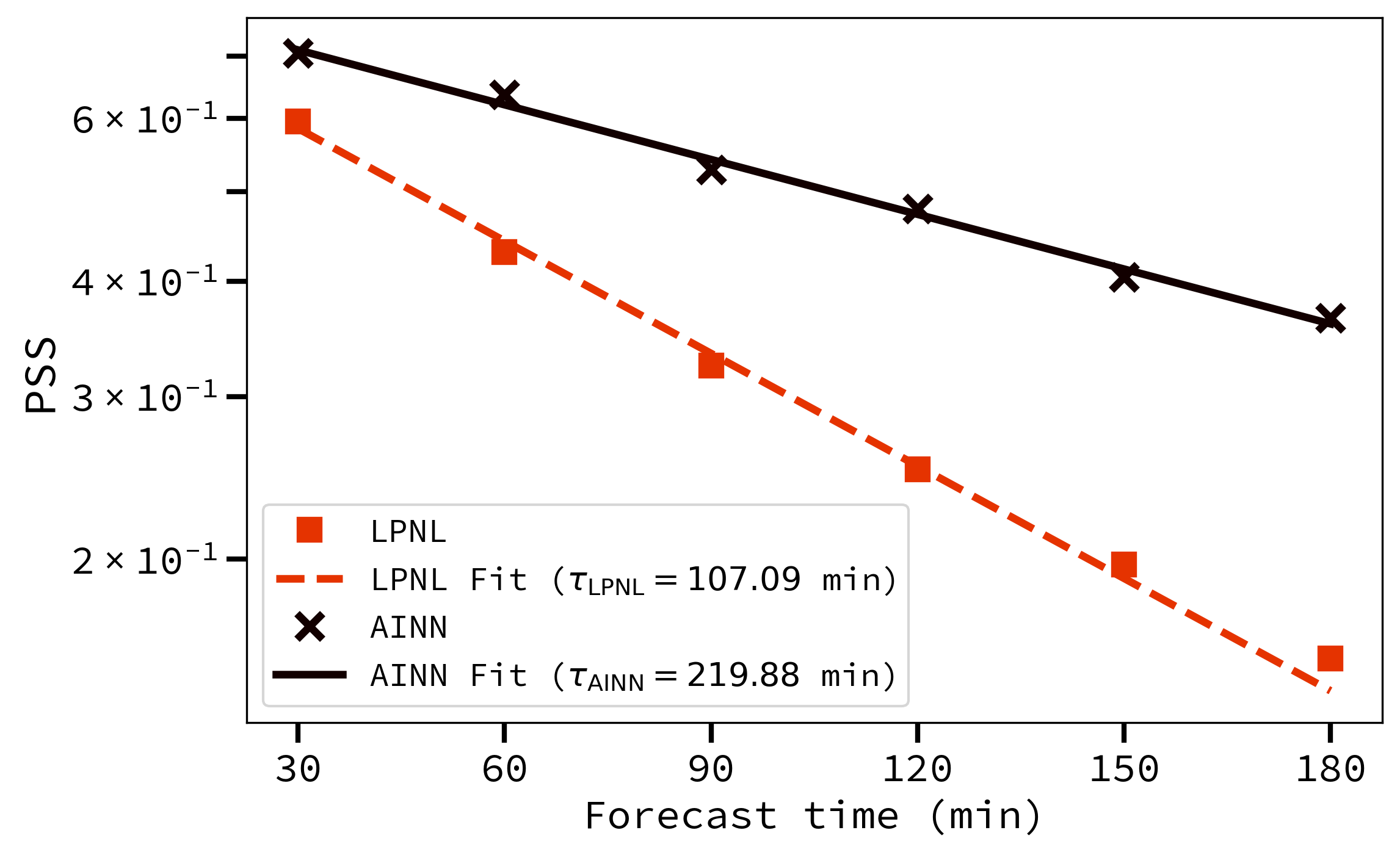}
\caption{Decay in skill with forecast time. The skill is expressed in terms of the PSS on a logarithmic vertical axis against the forecast time in minutes on the horizontal axis. The AINN is depicted with black crosses while the LPNL is represented by red boxes. The exponential fit for the AINN is shown as the solid black line and for the LPNL as a dashed red line.}
\label{fig:ComparisonAgainstLp}
\end{figure}

The AINN outperforms the LPNL on all scores considered in this work. Notably, the out-performance on all members of the HFB triplet implies that the AINN is superior to the LPNL in all aspects: The higher hit rate H signals a better ability to detect thunderstorms when they do occur, while the lower false alarm rate $F$ indicates less missed warnings. The bias B of the AINN is also closer to the perfect score of $1$,  meaning the frequency of predicted thunderstorms is closer to the frequency observed in the data. However, both models over-predict the occurrence of thunderstorms. The significantly higher CSI corroborates the notion that the skill improvement is not just due to more TN. The AINN achieves an overall \SI{60}{\percent} improvement over the LPNL in PSS. \cref{fig:ComparisonAgainstLp} displays the forecast time dependence of the PSS for both models.

 We detect an exponential decay in skill for both models, which is the typical and expected form for most forecasting skills \citep{selz_transition_2022, yousefnia_inferring_2024}. The time scales of the skill decay are quantified by fitting exponential functions $\exp{(-t/\tau_m)}$ for which one observes approximately the double value for the AINN $\tau_\text{AINN} = \SI{220}{\minute}$ compared to the LPNL $\tau_\text{LPNL} = \SI{107}{\minute}$. Moreover, the gap in skill between the AINN and the LPNL is significantly smaller for the shortest forecast time of \SI{30}{\minute} but widens with increasing forecast time, supporting the notion that for short forecast times the advection is the leading order process but for longer forecast times the life-cycle component gains importance.

 \subsection{Probabilistic Skill Metrics}\label{sec:skillMetrics}
\begin{figure}[htbp]
\centering
\includegraphics[width=\columnwidth]{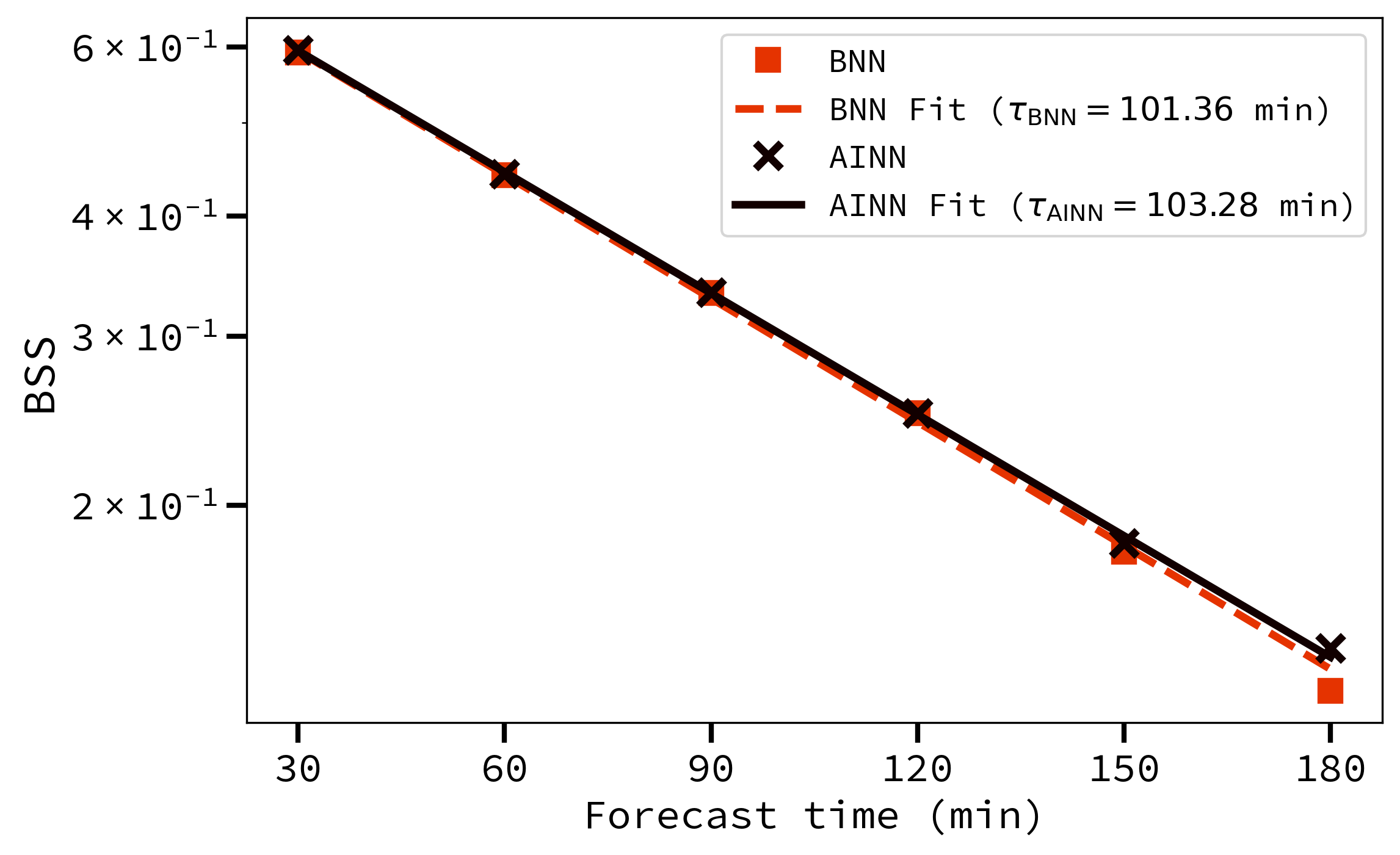}
\caption{Decay in skill with forecast time. The skill is expressed in terms of the BSS on a logarithmic vertical axis against the forecast time in minutes on the horizontal axis. The AINN is depicted with black crosses while the BNN is represented by red boxes. The exponential fit for the AINN is shown as the solid black line and for the BNN as the dashed red line.}
\label{fig:BssAgainstLeadtime}
\end{figure}

To compare the BNN to the AINN, we first employ the BSS. Averaged over all forecast times, the BSS for the AINN is $\text{BSS}_{\text{AINN}}=0.324$ compared to $\text{BSS}_{\text{BNN}}=0.320$. This constitutes a relative improvement of \SI{1.25}{\percent}. \cref{fig:BssAgainstLeadtime} displays the BSS and the exponential fits for both models against forecast time. The skill difference of the models stems entirely from the long forecast times \SI{150}{\minute} and \SI{180}{\minute} with \SI{2.2}{\percent} and \SI{10.8}{\percent} relative improvement respectively. This aligns with the prediction from our scale argument made in \cref{sec:receptField}. For a forecast time of \SI{180}{\minute} the scale argument presupposes a large scale advection speed of $v \gtrsim \SI{20}{\meter \per \second}$ for the advection nowcast to matter, which is still a high value but common enough to also notice a difference for the BSS calculated on climatologically consistent test sets. Furthermore, we again detect an exponential decay in skill with forecast time. The time scale of the AINN derived from the BSS is very different from the one derived from the PSS before. This is not surprising as there is no reason why the time scale for two scores should be the same. The time scales of the AINN and BNN do not differ substantially due to the performance difference showing up only for long forecast times.

\begin{figure}[htbp]
\centering
\includegraphics[width=\columnwidth]{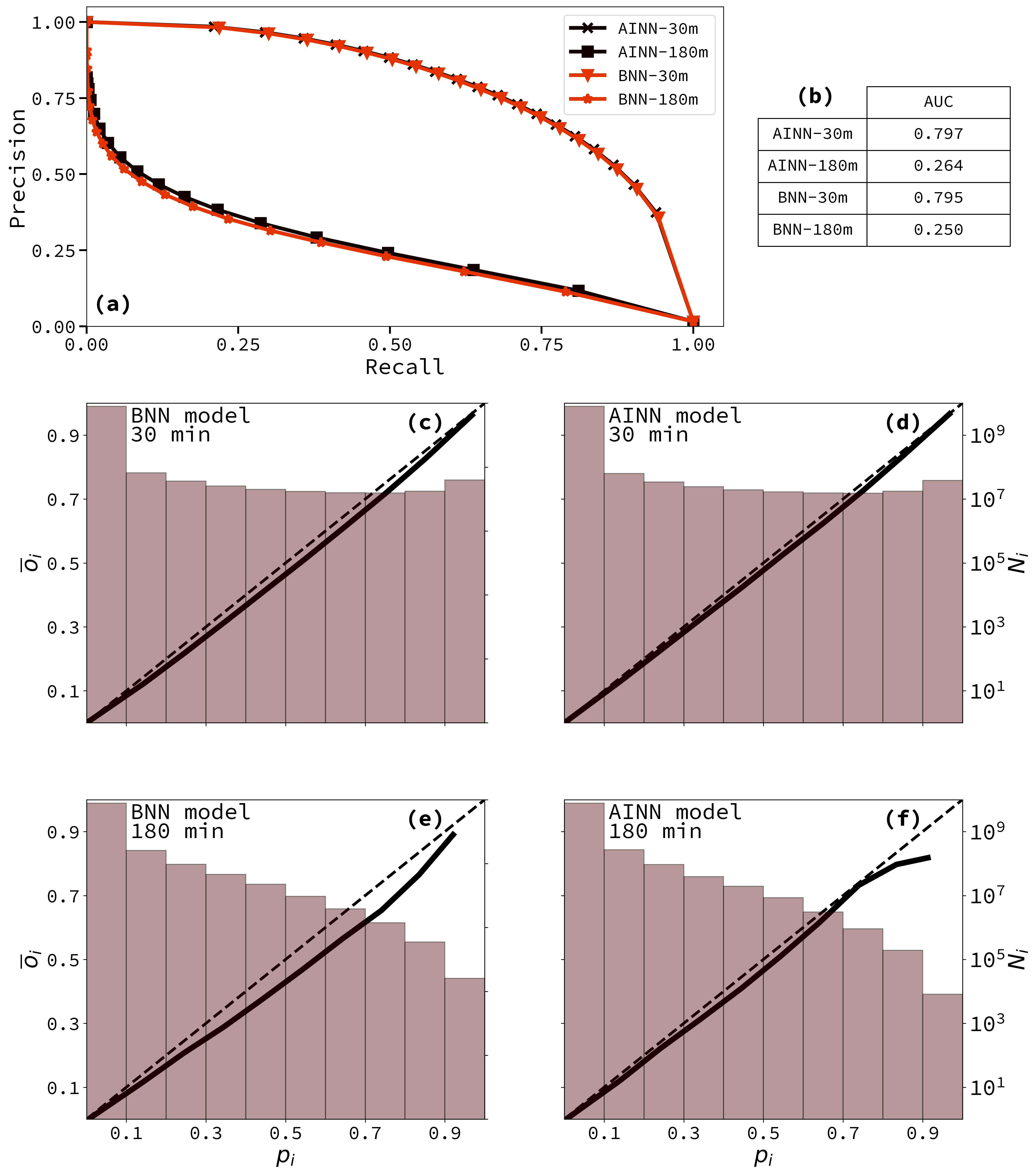}
\caption{(a) PR curves for forecast times of \SI{30}{\minute} and \SI{180}{\minute}. The curves of the AINN are in black and of the BNN in red. (b) Table detailing the AUC scores. Reliability diagrams for the BNN for forecast times of \SI{30}{\minute} in (c) and \SI{180}{\minute} (e) and the AINN for forecast times of \SI{30}{\minute} in (d) and \SI{180}{\minute} (f). The number of bins was chosen $N_\text{b}=10$. The solid black line is the plot of the observed event frequency $\overline{o}_i$ against the bin-averaged forecast probability $p_i$. The dashed black diagonal line constitutes the reference for perfect calibration. The refinement distribution is displayed as the red histogram.}
\label{fig:ReliabilityDiagrams}
\end{figure}

Further tools for evaluating the models are the reliability diagram and the PR curve. They are compared for a short forecast time of \SI{30}{\minute} and a long forecast time of \SI{180}{\minute} in \cref{fig:ReliabilityDiagrams}. The reliability diagrams imply that both the BNN and the AINN are well calibrated for a forecast time of \SI{30}{\minute}, with the AINN displaying a slight edge. The refinement distribution of both models suggests many correct high-confidence predictions close to \SI{0}{\percent} and \SI{100}{\percent}. However, for a forecast time of \SI{180}{\minute}, the calibration gets worse for both models and they display distinct phenomenology. The BNN is overconfident for probabilities between \SI{30}{\percent} and \SI{85}{\percent} while being well calibrated otherwise. The AINN, on the other hand, is well calibrated for all probabilities up to around \SI{80}{\percent}, above which it becomes increasingly overconfident. The access to the Lagrangian persistence nowcasts of the input channels enables the model  to course correct for a large probability region but also incites some overconfidence for very high probabilities. The refinement distribution further underscores the difficulty of both models to produce correct high-confidence predictions for long forecast times. From the PR curves one observes that for a forecast time of \SI{30}{\minute} the curves of the AINN and the BNN  essentially overlap, which is also reflected by their AUC scores differing only on the third digit. For a forecast time of \SI{180}{\minute}, the AINN curve slightly outperforms the BNN on medium thresholds, which leads to a relative improvement in the AUC score of \SI{5.6}{\percent}. The observed strong drop in skill as forecast time increases, the lack of difference between the models for short forecast times, and the noticeable improvement of the AINN over the BNN for long forecast times as measured by the AUC, are in full agreement with the BSS-based assessment and the predictions from the scale argument.

\subsection{Skill Metrics Conditioned on Advection Speed and forecast time}\label{sec:skillAdvectionSpeed}
\begin{figure}[htbp]
\centering
\includegraphics[width=\columnwidth]{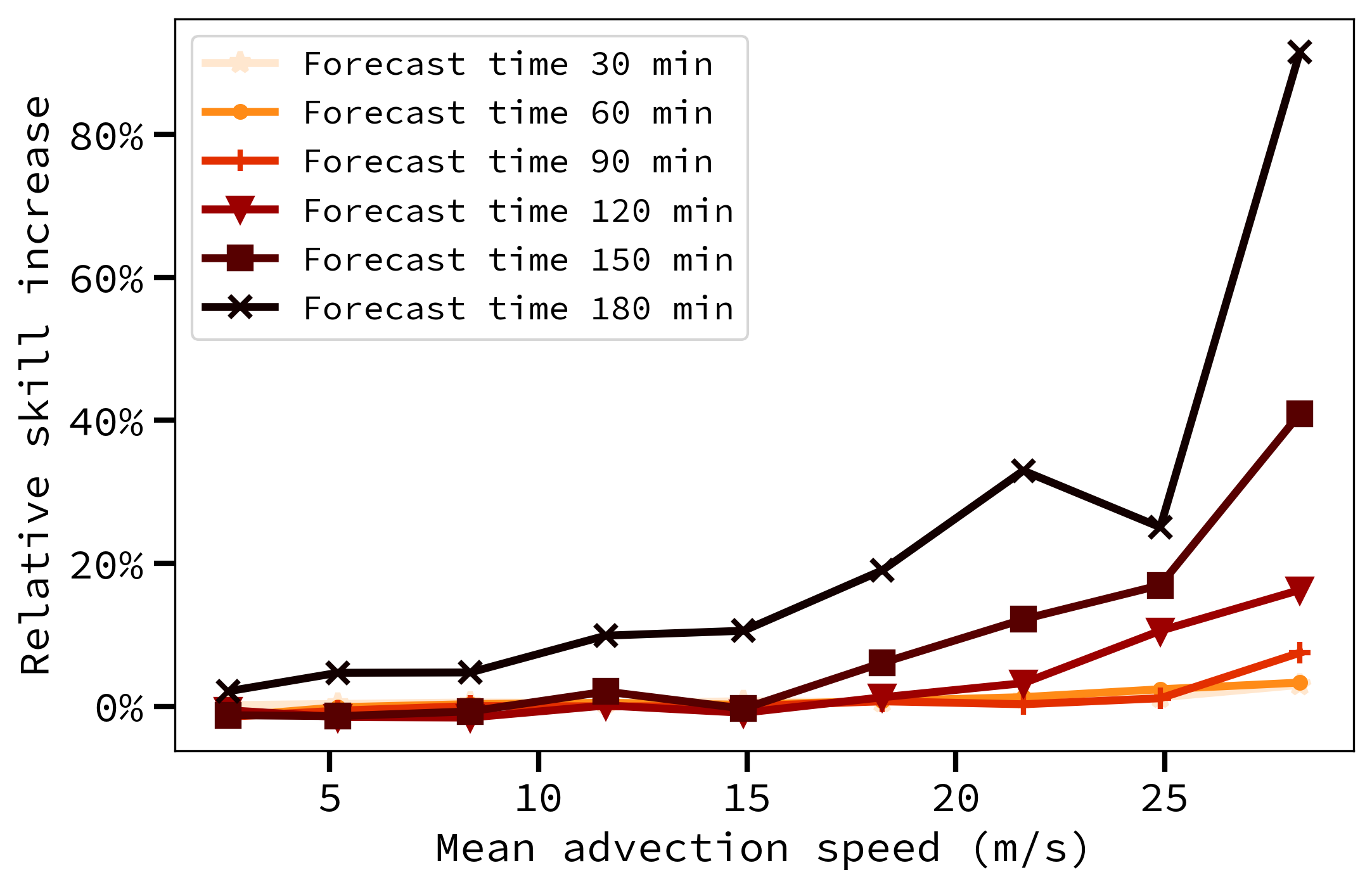}
\caption{Relative skill increase of the AINN over the BNN (in percent) against the advection speed (in meters per second) for all forecast times considered in this study. The original skill is measured with the BSS. In isolated cases — particularly at low advection speeds (\SI{0}{–}\SI{15}{\meter\per\second}) and forecast times up to 150 minutes — the AINN slightly underperforms the BNN, with a maximum relative skill decrease of less than 1.5\%.}
\label{fig:SkillAgainstAdvectionSpeed}
\end{figure}

To test the validity of our scale argument more precisely, we focus on its implication that the relative skill increase of the AINN over the BNN should increase with forecast time and advection speed. In the previous section, we conditioned the skill purely on the forecast time. Now we also condition the skill on the advection speed as follows: For each forecast time, we calculate for each crop in the test set $D_{t,m}(\text{test})$ the average of the advection speed over the advection field derived from the WV062 channel. We further partition the advection speed ranging from (\SI{0}{\meter \per \second}, \SI{30}{\meter \per \second}) into $9$ bins and assign each data point to its bin accordingly. Data points with average advection speed above \SI{30}{\meter \per \second}, which are less than \SI{8}{\percent} of the dataset, are purposefully ignored. These extreme values span a nonphysically wide range (up to \SI{678}{\meter \per \second}) and are likely artifacts of the optical flow algorithm failing in clear-sky conditions, where no coherent motion can be detected. This interpretation is supported by the very low lightning occurrence (0.24\%) in this regime. Then we calculate the average BSS for both the baseline and the AINN per bin. Importantly, we utilize for the calculation of the reference BS for each bin the specific climatology of that bin, as thunderstorms are not uniformly distributed across the bins. In a final step, the relative skill improvement of the AINN over the BNN is calculated for each bin. The final result is displayed in \cref{fig:SkillAgainstAdvectionSpeed}, where the bin-wise relative improvement in skill is plotted against the mean advection speed value of each bin for all forecast times. 


There is no relative skill improvement for short forecast times, except for very high advection speeds and even then it is very minor. For longer forecast times one observes a significant improvement for high advection speeds. There is a clear trend emerging where the relative improvement in skill increases with advection speed and forecast time. We conclude that the AINN provides a worthwhile benefit over the baseline but only under specific circumstances, namely long forecast times and high advection speeds. The explanation in terms of relating the underlying physical scales to the intrinsic scale of the neural networks is supported by this analysis.

Two limitations of the study should be mentioned here: First, this analysis and also the scale argument in general suppose that the large-scale advection as estimated from only two observations with a \SI{15}{\minute} time interval is close to correct and should be approximately constant over the forecast period of \SI{3}{\hour}. To partly investigate the stability of the results against this point, additional experiments are performed with 4 input time steps for a forecast time of \SI{180}{\minute}. The results are detailed in \cref{sec: additional experiments}. We find that additional input time steps do not change the AINN but moderately improve the BNN performance. While this decreases the skill difference, it does not change our findings on a fundamental level. Moreover, the receptive field calculation is an estimate in the sense that it provides the maximum side length of the box of input pixels that the network can make use of for a prediction. But this does not mean the network has to make use of all those pixels. In many computer vision applications, the effective receptive field is significantly smaller than the theoretically calculated maximum \citep{luo_understanding_2016}.

\subsection{Case Study}\label{sec:caseStudy}
\begin{figure}[htbp]
\centering
\includegraphics[width=\columnwidth]{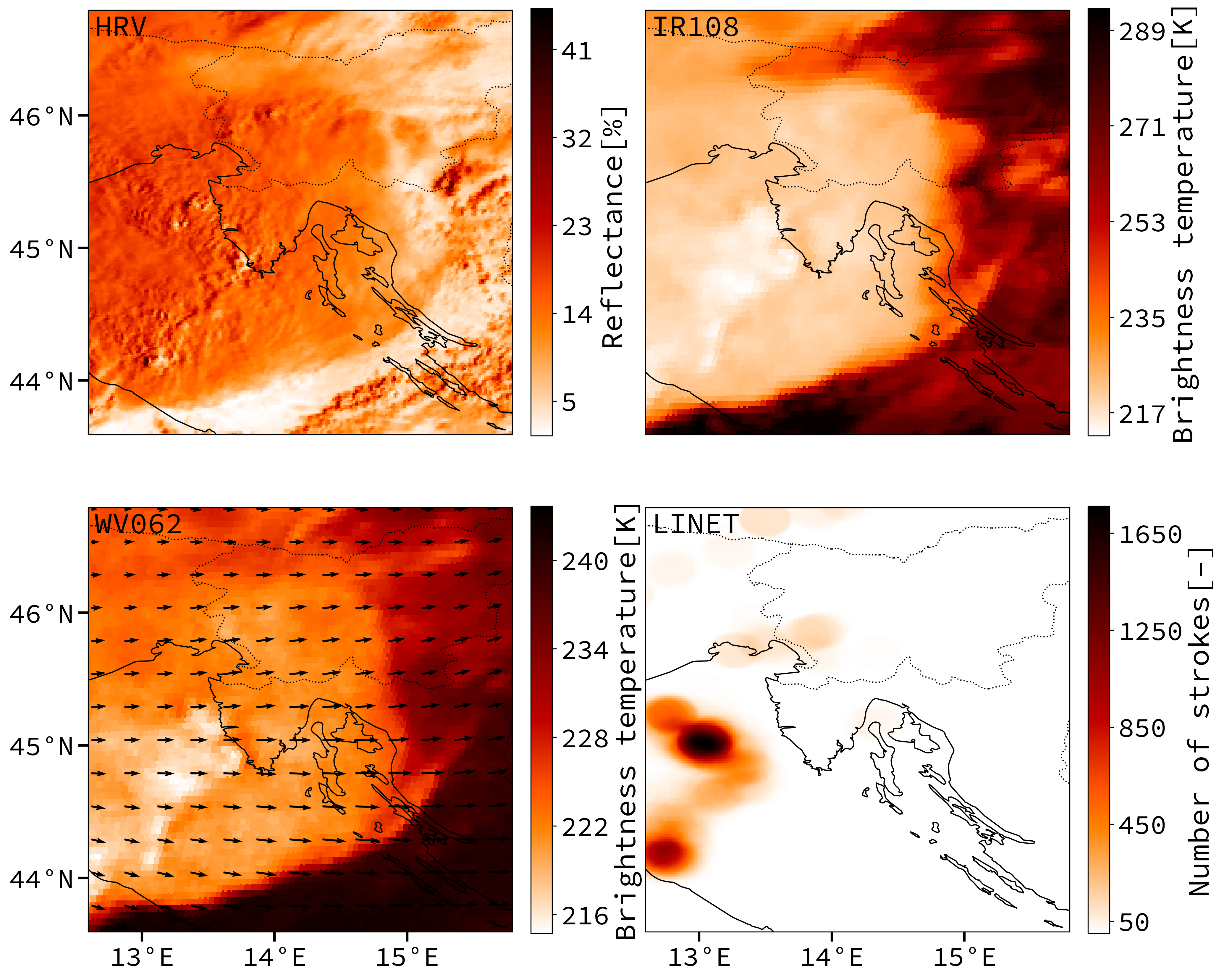}
\caption{Case study on August 2, 2019, 1700~UTC over the Adriatic sea and Croatia to illustrate the idea behind the scale argument. All observations are at the time of the input. The top left panel constitutes the HRV channel, while the top right panel represents the IR108 channel. The bottom left panel is the satellite imagery of the WV062 channel with its advection field as an overlay. The bottom right panel displays the aggregated LINET observations.}
\label{fig:InputCaseStudy}
\end{figure}

We close with a discussion of a case study from the test set, which reinforces the mechanism proposed in the scale argument. The starting point is on August 2, 2019, 1700~UTC from which a \SI{2}{\hour} nowcast is performed for 1900~UTC. The weather situation is inferred from \cref{fig:InputCaseStudy}: Satellite imagery indicates a well-developed thunderstorm complex situated over the Adriatic sea close to the coast of Croatia. In the HRV channel, dense cloud shields and overshooting tops are apparent, reflecting strong convective updrafts. The IR108 imagery shows cloud-top temperatures dropping below \SI{220}{\kelvin} in the most intense cells, highlighting very cold, high-reaching cloud tops typically associated with severe thunderstorms. Lightning stroke density is elevated, confirming vigorous electrical activity within the storm system. A strong westerly flow with an average advection direction of \SI{267.35}{\degree} and advection speed of \SI{22.20}{\meter \per \second} (both measured from the WV062 channel) is steering the convection eastward towards the interior of Croatia. Given the ongoing robust convection and continued forcing from the strong westerly flow, these storms are likely to persist and track further east over the next two hours.

\begin{figure}[htbp]
\centering
\includegraphics[width=\columnwidth]{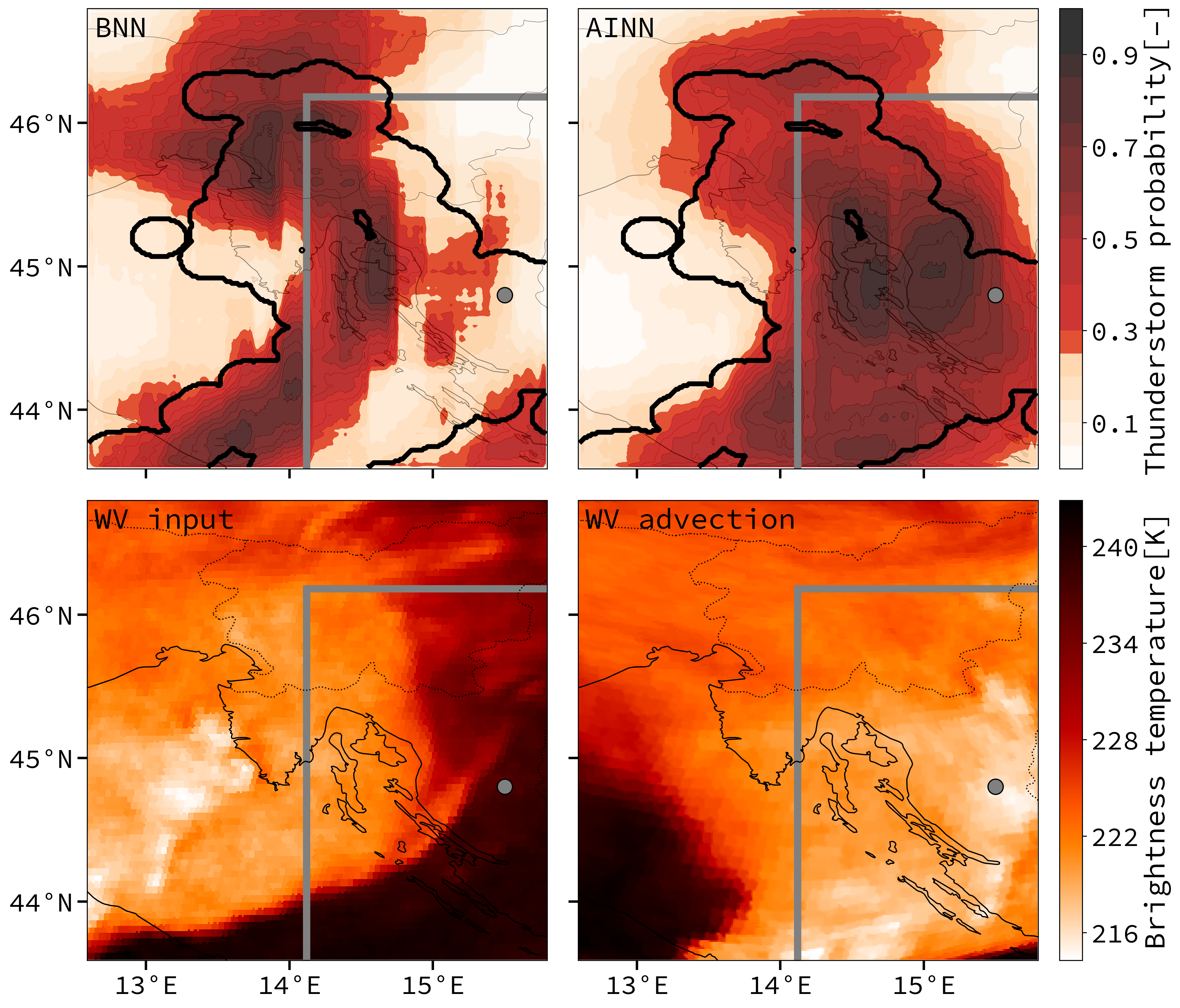}
\caption{Case study on August 2, 2019, 1700~UTC over the Adriatic sea and Croatia to illustrate the idea behind the scale argument. The top left panel constitutes the \SI{2}{\hour} prediction of the BNN, while the top right panel represents the prediction of the AINN. The model output is thresholded using the best-CSI threshold such that red shading is counted as a thunderstorm prediction. The black contours indicate the ground truth thunderstorm activity as measured by lightning observations. The bottom left panel is the satellite imagery of the WV062 channel at the time of the input and the bottom right panel its Lagrangian persistence nowcast for the desired \SI{2}{\hour} mark. The gray dot constitutes a pixel that is correctly forecasted as a thunderstorm pixel by the AINN but not by the BNN. The gray box indicates the boundary of that pixel's receptive field.}
\label{fig:CaseStudy}
\end{figure}

\cref{fig:CaseStudy} displays observations, predictions and an example of a receptive field relevant to this case. We observe that the BNN does roughly capture the western part of the storm but fails to nowcast the eastern part. The AINN on the other hand is capable of correctly predicting this part as well. We argue that the reason for this lies in the receptive field and the additional Lagrangian persistence information available to the AINN. The most intense part of the storm (brightness temperature below \SI{217}{\kelvin}) lies outside of the receptive field (gray box is the boundary) of the gray point highlighted in \cref{fig:CaseStudy}. Thus, the BNN is not aware of that storm part and cannot take it into account for its nowcast of the eastern part. This is also initially the case for the AINN. However, it also has access to the Lagrangian persistence nowcast of the input channels. Therefore, the intense part of the storm is advected into the receptive field of the eastern region and can be considered. We also note that there exists a small number of points for which the most intense part of the storm is at the very edge of their receptive fields; yet, the BNN is not capable of confidently predicting a thunderstorm. This could be a case of the effective receptive field being somewhat smaller than the theoretical maximum.
We conclude this section with the insight that for such a situation as described in the case study, it requires long forecast times and high advection speeds as presupposed by the scale argument.

\section{Discussion}\label{sec:discussion}
This study investigated the generality of combining classical advection algorithms with CNNs for nowcasting by applying it to the novel context of satellite-based thunderstorm nowcasting. For each of the two considered model classes a separate ResU-Net was trained for each forecast time to solve a binary segmentation task, where lightning observations served as ground truth. The BNN utilized the latest two time steps of four satellite channels and aggregated lightning observations to predict a pixel-wise probability of thunderstorm occurrence, while the AINN additionally received the Lagrangian persistence nowcast of each input channel.

Furthermore, we proposed an explanation of why and under what circumstances the AINN should outperform the BNN. This was posited as a scale argument relating the underlying physical scales (forecast time and large-scale average advection speed) to the intrinsic scale of the CNNs (the side length of the receptive field). The main implications of the scale argument were the prediction that the AINN should outperform the BNN considerably only for forecast times larger than \SI{2}{\hour} and that the relative improvement in skill should increase both with forecast time and advection speed.

The AINN outperformed---as measured by the BSS---the BNN as predicted for forecast times greater than \SI{2}{\hour}. For a forecast time of \SI{3}{\hour}, the skill improved by \SI{10.8}{\percent} through the inclusion of the advection nowcasts. This was further supported by the analysis of reliability diagrams, where one, however, also observed that for long forecast times the AINN becomes slightly overconfident for high probability outputs. 

To test the implication of the scale argument of a stronger effect with higher advection speeds and forecast time, we analyzed the relative improvement in skill of the AINN over the BNN conditioned both on forecast time and large-scale mean advection speed. We concluded that the predicted trend of an increase with forecast time and advection speed was clearly present. 

Finally, the scale argument was illustrated by a case study which further underscored that the usefulness of the AINN indeed stems from cases where the advection transports relevant information through high advection speeds and long forecast times into the receptive field of the network.

This study did not explore alternative approaches to circumvent undersized receptive fields. A thorough investigation of adding more layers or incorporating dilated convolutions is left to future work. Another promising direction would be to explore transformer-based architectures, which have recently become standard in many vision and segmentation tasks. Due to their self-attention mechanism, transformers can naturally incorporate long-range spatial dependencies.

We close by emphasizing the general usefulness of this work. CNNs are ubiquitous in spatiotemporal forecasting problems and are often applied to problems where some prior domain knowledge exists. In such cases, scale arguments relating the underlying scales of the problem to the inherent scales of the neural networks in the shape of the receptive field could be a valuable tool to reach optimal architecture and hyperparameter choices without spending a large compute budget on systematically testing a lot of reasonable combinations.

\clearpage
\acknowledgments
We gratefully acknowledge the computational and data resources provided through the joint high-performance data analytics (HPDA) project "terrabyte" of the DLR and the Leibniz Supercomputing Center (LRZ). C.M. carried out his contributions within the Italia--Deutschland Science--4--Services Network in Weather and Climate (IDEA-S4S; SESTO, 4823IDEAP4). This Italian-German research network of universities, research institutes and DWD is funded by the Federal Ministry of Digital and Transport (BMDV). The authors declare that there are no conflicts of interest to disclose. 

%
%
\datastatement
The MSG SEVIRI data are available to EUMETSAT members and participating organizations at the EUMETSAT Data Store (\url{https://data.eumetsat.int/}). The lightning data are proprietary and can be purchased from the nowcast GmbH (\url{https://www.nowcast.de/en/}).

%

\appendix




\appendixtitle{Supplementary information}

%

\subsection{Hyperparameter Specifications}

\cref{tab:training_hyperparameter} summarizes all chosen hyperparameters and their tested variations of this study. No variations to the number of epochs were tested, as all models converged at this point. All variations were tested only for the BNN with forecast time of $t=$\SI{30}{\minute}. No comprehensive hyperparameter grid sweep was performed. Instead, variations were tested sequentially in the order listed in the table: for each parameter, all its variations were evaluated, the optimal value was selected and fixed, and then the next parameter was tuned in the same manner. Most variations did not strongly affect results, so further tuning by model class or forecast time was assumed unlikely to yield significant improvements.

\begin{table}[htbp]
\caption{Summary of training specific hyperparameter and tested alternative variations. If a parameter is not listed, it was kept at the pytorch default value and no variations for it were tested. The parameters factor, patience and cooldown are specific to the ReduceLROnPlateau learning rate scheduler.}
\label{tab:training_hyperparameter}
\centering
\begin{tabular}{ccc}
\toprule
Parameter &  Chosen Value & Tested variations \\
\midrule
Number epochs & $60$ & - \\
Batch size & $128$  & $32, 64$  \\
L2 regularization & $10^{-4}$ & $10^{-3}, 10^{-5}$ \\
Initial learning rate & $10^{-3}$ & $10^{-2}, 10^{-4}$ \\
Factor & $0.1$ & - \\
Patience & $5$ & $3$ \\
Cooldown & $3$ & $2$ \\
\bottomrule
\end{tabular}
\end{table}

\subsection{Thresholds to maximize CSI}

\cref{tab:thresholds_CSI} lists the thresholds that maximize the CSI which are then used to calculate the categorical skill scores listed in \cref{tab:skill_scores}.

\begin{table}[htbp]
\caption{Summary of probability thresholds that maximize CSI dependent on forecast time and model class.}
\label{tab:thresholds_CSI}
\centering
\begin{tabular}{ccc}
\toprule
Forecast time &  AINN & BNN \\
\midrule
\SI{30}{\minute} & $0.4$ & $ 0.4$ \\
\SI{60}{\minute} & $0.3$  & $0.35$  \\
\SI{90}{\minute} & $0.3$ & $0.3$ \\
\SI{120}{\minute} & $0.25$ & $0.25$ \\
\SI{150}{\minute} & $0.25$ & $0.2$ \\
\SI{180}{\minute} & $0.2$ & $0.2$ \\
\bottomrule
\end{tabular}
\end{table}

\subsection{Additional experiments on the number of time steps} \label{sec: additional experiments}

To probe the impact of the number of time steps $T_m$ on the results, we perform additional experiments. We retrain both models for a forecast time of \SI{180}{\minute}. The longest forecast time was specifically chosen because the largest benefit for including the advection was observed for it. Furthermore, one would expect a longer input time series to improve the advection for long forecast times the most. The baseline neural network now receives the last $4$ observations as input and is referred to as BNN-4. Additionally to these $4$ time steps, the advection-informed neural network also receives the Lagrangian persistence nowcast at the forecast time. It is therefore referred to as AINN-5. In short this means $T_{\text{BNN-4}}=4$ and $T_{\text{AINN-5}}=5$. To clearly distinguish between the four models with forecast time of \SI{180}{\minute}, we also rename the models specified in \cref{sec:NN} that are based on two input time steps to BNN-2 and AINN-3. We want to stress that we also recalculate the advection fields and the Lagrangian persistence nowcasts based on four input time steps to arrive at the AINN-5 model. For that we utilize the capability of the pySTEPS implementation of the Lucas-Kanade algorithm to temporally average over multiple time steps. 

\cref{tab:skillScoresAppendix} compares the BSS and the AUC of the models. First, we observe that there is no improvement of the AINN-5 over the AINN-3 in terms of the BSS, while the AUC gets worse. From that, we tentatively conclude that additional time steps and a temporally-averaged advection field does not improve an advection informed model. We did, however, not explore if the additional model capacity in form of a larger number of trainable parameters changes this finding. Second, the BNN-4 shows improvement over the BNN-2 in terms of BSS, while the AUC gets worse, but less than for the advection informed models. Therefore, we observe that the relative improvement of the AINN-3 over the BNN-2 of $10.8\%$ in terms of the BSS is reduced to a $7.6\%$ improvement of the AINN-5 over the BNN-4. Similarly, the AINN-3 improves upon the BNN-2 in terms of the AUC by $5.6\%$, while the improvements of the AINN-5 over the BNN-4 is reduced to $4.4\%$. 

\begin{table}[htbp]
\caption{Skill scores for all models with forecast time of \SI{180}{\minute}.}
\label{tab:skillScoresAppendix}
\centering
\begin{tabular}{ccc}
\toprule
Model &  BSS & AUC \\
\midrule
AINN-3   & $0.142$ & $0.264$ \\
AINN-5 & $0.142$  & $0.260$  \\
BNN-2 & $0.128$ & $0.250$ \\
BNN-4  & $0.132$ & $0.249$ \\
\bottomrule
\end{tabular}
\end{table}

\begin{figure}[htbp]
\centering
\includegraphics[width=\columnwidth]{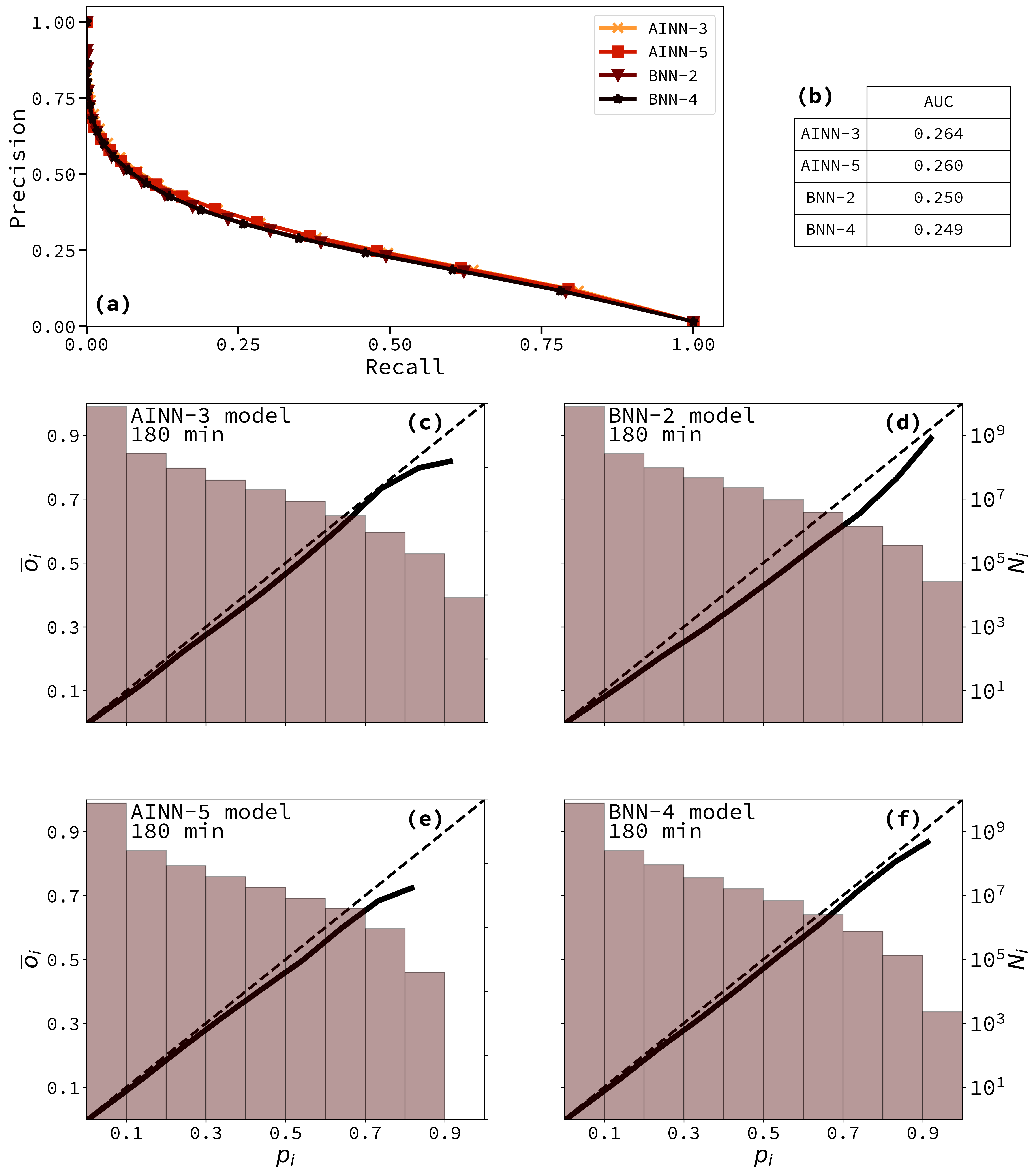}
\caption{(a) PR curves for forecast times of \SI{180}{\minute} for multiple time steps and model classes. (b) Table detailing the AUC scores. Reliability diagrams for models with  forecast times of \SI{180}{\minute}: (c) AINN-3, (d) BNN-2, (d) AINN-5, (f) BNN-4. The number of bins was chosen $N_\text{b}=10$. The solid black line is the plot of the observed event frequency $\overline{o}_i$ against the bin-averaged forecast probability $p_i$. The dashed black diagonal line constitutes the reference for perfect calibration. The refinement distribution is displayed as the red histogram.}
\label{fig:ReliabilityDiagramsAppendix}
\end{figure}

In \cref{fig:ReliabilityDiagramsAppendix} the PR curves of all models with forecast time of \SI{180}{\minute} together with their reliability diagrams are displayed. Again, we observe that the AINN-5 does not improve much compared to the AINN-3 model, while the BNN-4 model improves upon the BNN-2 in reliability.

\begin{figure}[htbp]
\centering
\includegraphics[width=\columnwidth]{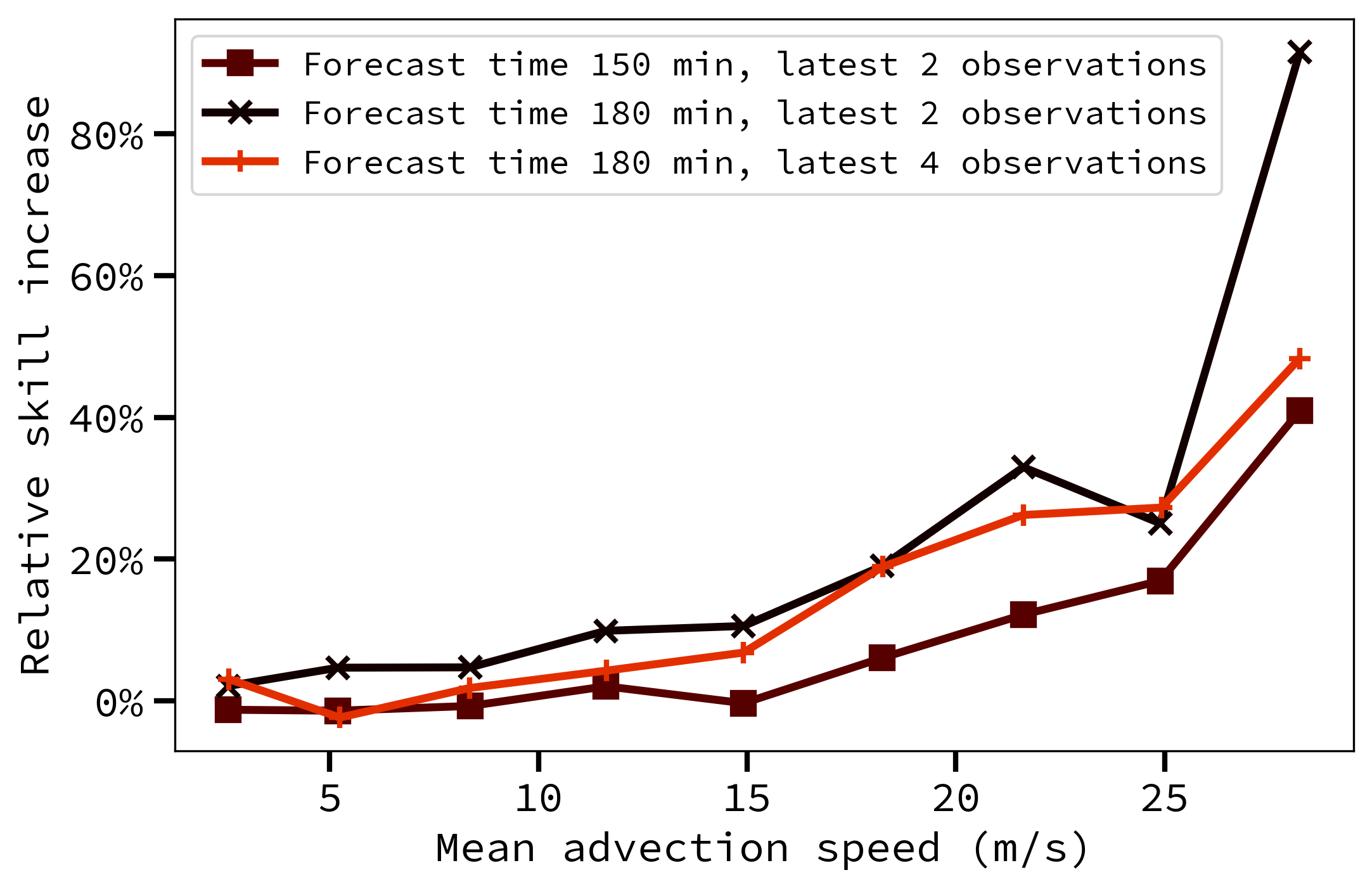}
\caption{Relative skill increase (in percent) of the advection informed models over the baseline models in terms of the BSS against the advection speed (in meters per second). AINN-5 over the BNN-4 for a forecast time of \SI{180}{\minute} in light red.   AINN-3 over the BNN-2 for a forecast time of \SI{180}{\minute} in black and for a forecast time of \SI{150}{\minute} in dark red.}
\label{fig:SkillAgainstAdvectionSpeedAppendix}
\end{figure}

We also redid the analysis of the skill improvement conditioned on the advection speed. The results are shown in \cref{fig:SkillAgainstAdvectionSpeedAppendix}. We find that the relative improvement of the AINN-5 over the BNN-4 is less than the relative improvement of the AINN-3 over the BNN-2 for most advection speed bins. The trend, however, is still in line with the prior findings.

To summarize: First, additional time steps do not benefit the advection-informed model (neither on the network input level, nor on the Lagrangian persistence estimation based on temporally-averaged advection fields). This is in line with our hypotheses: It is only important that the relevant parts of the observations are inside of the receptive field of the network, not so much that the estimation of the advection field in itself is perfect. The baseline model, however, improves with additional time steps. As it does not receive additional information, it can improve its estimation of the advection and close the gap on the advection-informed model. However, it still struggles with high advection speeds as the relevant information is still not available to the network due to its limited receptive field. We conclude that while additional time steps decrease the effect size, they do not change our findings on a fundamental level. The hypothesis in form of the scale argument is still supported by the experiments. 



\bibliographystyle{ametsocV6}
\bibliography{main.bbl}

\end{document}